\documentclass[10pt,oneside,twocolum,final]{IEEEtran}

\usepackage{amsmath,amsfonts}
\usepackage{amssymb}
\usepackage{algorithmic}
\usepackage{algorithm}
\usepackage{bm}
\usepackage{booktabs}
\usepackage{array}
\usepackage[caption=false,font=normalsize,labelfont=sf,textfont=sf]{subfig}
\usepackage{extpfeil}
\usepackage{textcomp}
\usepackage{stfloats}
\usepackage{url}
\usepackage{verbatim}
\usepackage{graphicx}
\usepackage{cite}
\usepackage{color}
\allowdisplaybreaks[4]

\begin{document}


\title{Exploring the Advantages of Sparse Arrays in Near-Field XL-MIMO Systems: Beam Analysis and EDoF Function}
\IEEEoverridecommandlockouts

\author{Xianzhe~Chen,
        Hong~Ren,~\emph{Member},~\emph{IEEE},
        Cunhua~Pan,~\emph{Senior~Member},~\emph{IEEE},
        Cheng-Xiang~Wang,~\emph{Fellow},~IEEE,
        and~Jiangzhou~Wang,~\emph{Fellow},~IEEE

\thanks{\emph{(Corresponding author: Cunhua Pan)}}

\thanks{X. Chen, H. Ren, C. Pan and Jiangzhou Wang are with the National Mobile Communications Research Laboratory, School of Information Science and Engineering, Southeast University, Nanjing 211189, China (e-mail: chen.xianzhe, hren, cpan, j.z.wang@seu.edu.cn).}

\thanks{Cheng-Xiang Wang is with the National Mobile Communications Research Laboratory, School of Information Science and Engineering, Southeast University, Nanjing 211189, China, and also with the Pervasive Communication Research Center, Purple Mountain Laboratories, Nanjing 211111, China (e-mail: chxwang@seu.edu.cn).}



}

\maketitle

\newtheorem{lemma}{Lemma}
\newtheorem{theorem}{Theorem}
\newtheorem{remark}{Remark}
\newtheorem{corollary}{Corollary}
\newtheorem{proposition}{Proposition}

\begin{abstract}
This paper investigates near-field XL-MIMO systems with sparse uniform planar arrays (UPAs).
Based on the Green's function-based channel model, 
the paper
derives closed-form expressions for the signal beam power when the distance coordinate or the angular coordinates varies with respective to the focused position.
Based on that, closed-form expressions for the lobe length and the suppressing ratio are obtained,
indicating that both the distance-focusing property and the grating lobe behavior can be enhanced as the focal distance decreases or the antenna spacing increases.
Furthermore, the paper introduces a crucial constraint on system parameters, under which effective degrees-of-freedom (EDoF) of XL-MIMO systems with sparse UPAs can be precisely estimated. 
Then, the paper proposes an algorithm to obtain a closed-form expression, which can estimate EDoF with high accuracy and low computational complexity.
The numerical results verifies the correctness of the main results of this paper.


\end{abstract}

\section{Introduction}\label{SA_Section1}

Massive multiple-input multiple-output (MIMO) has been a key technology in fifth-generation (5G) communications for its merits in terms of spectral efficiency, energy efficiency and power control \cite{6375940,6457363,6736761,Fundamental_2016,8861014}.
However, future sixth-generation (6G) communications
set higher requirements, including ultra-reliability, high capacity densities, extremely low-latency and low-energy consumption.
To meet these requirements, 
extremely large-scale MIMO (XL-MIMO),
as an advanced evolution of massive MIMO, 
has garnered significant attention in recent researches \cite{BJORNSON20193,8766143,10054381,10098681,10496996,10536068}.
Compared to conventional massive MIMO, 
XL-MIMO employs an order-of-magnitude larger number of antennas 
to achieve exceptionally high spectral efficiency. The significantly increased number of antennas 
not only expands the array size 
but also shifts the system's operational environment from the traditional far-field region to the near-field region.
As a result, new channel characteristics appears such as the spherical wavefront, spatial non-stationarity and so on \cite{9170651,9617121,10500425,10646401}.

An important benefit of spherical wavefronts in near-field XL-MIMO systems is the ability for distance focusing. It guarantees that the power of the signal beam sent by XL-MIMO arrays can be concentrated on the range near the focused position along the distance direction \cite{10443535,Towards6G}. This offers a significant advantage for multi-user systems, as users can be distinguished by their distance coordinates, which is fundamentally different from the far-field scenario that relies exclusively on angular separation.

However, most existing studies on XL-MIMO systems assume half-wavelength antenna spacing, resulting in high power consumption and hardware costs due to the large number of antenna elements. To overcome this limitation, sparse arrays have been proposed as a promising alternative, utilizing a significantly smaller number of antennas within a given aperture. In \cite{10545312}, an XL-MIMO system based on modular sparse uniform linear arrays (ULAs) was investigated, demonstrating the distance-focusing property of such configurations. This phenomenon was further illustrated in \cite{10496996}. Additionally, the authors of \cite{11039147} analyzed the beam pattern of sparse ULAs and revealed a constrained focusing depth in the range dimension.

\begin{figure*}[b]
	\centering
	\includegraphics[scale=0.65]{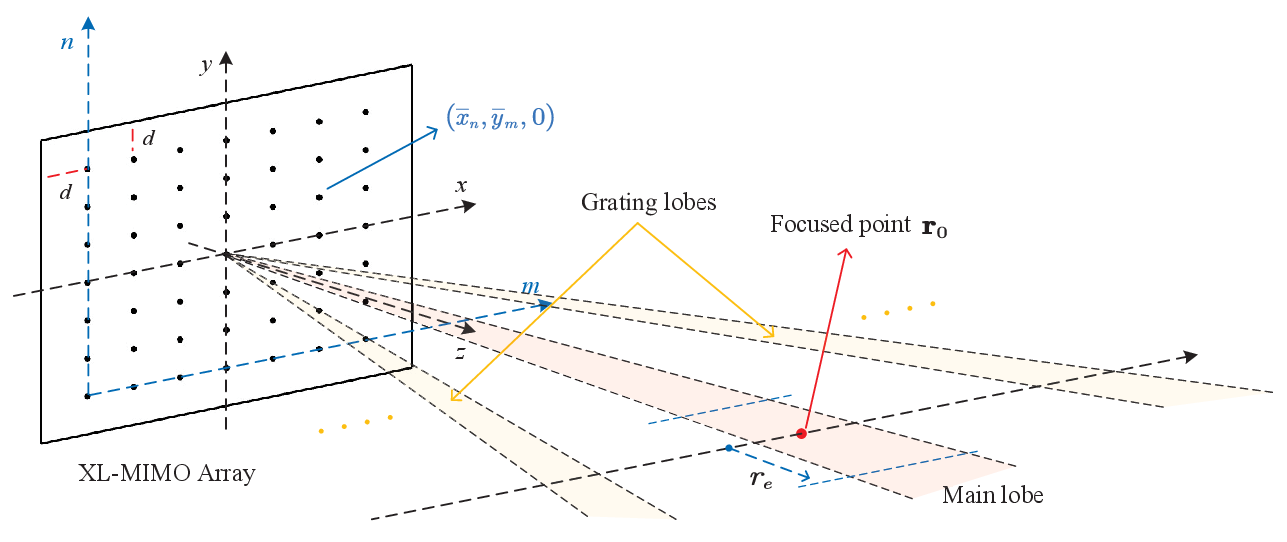}\\  
	\caption{System model for an XL-MIMO system with a sparse UPA.}\label{SA_p2} 
\end{figure*}

It is important to note that sparse arrays tend to generate unintended grating lobes, whose amplitudes are comparable to that of the main lobe in the far field. These grating lobes introduce substantial inter-user interference, presenting a major challenge in conventional far-field multi-user communication systems. To mitigate this issue, effective grating lobe suppression techniques are essential when deploying sparse arrays in far-field scenarios \cite{1229889,5613283,9898900}.
In contrast, the work in this paper reveals that in near-field XL-MIMO systems using sparse arrays, the grating lobes are naturally suppressed, especially when the antenna spacing is increased. This suppression leads to a significant reduction in grating lobe-induced interference, thereby supporting the practical feasibility of sparse arrays for near-field XL-MIMO applications.

An additional advantage offered by the spherical wavefront in near-field XL-MIMO systems is the increase in effective degrees of freedom (EDoF), defined as the number of significant singular values of the channel matrix. This improvement becomes even more pronounced when sparse arrays are employed \cite{chen2025}. Consequently, accurately estimating the EDoF in such systems is of considerable importance.
Several approaches for EDoF estimation have been proposed in the literature. For instance, a concise estimation method based on the maximum number of intensity fringes was introduced in \cite{2019Waves}. Using two-dimensional sampling theory, approximate expressions for the EDoF between a large intelligent surface (LIS) and a small intelligent surface (SIS) were derived in \cite{9139337}. The impact of EDoF on channel capacity in systems with two non-parallel arrays was analyzed in \cite{9860745}. In the context of free-space MIMO systems, \cite{9650519} investigated the EDoF through direct computation from the channel matrix. Moreover, \cite{10856805} presented closed-form expressions for the EDoF in near-field XL-MIMO systems using Green’s function-based channel models.
It should be noted, however, that existing EDoF estimation methods often suffer from either limited accuracy or high computational complexity. Therefore, there is a clear need for a method that can achieve high estimation accuracy while maintaining low computational cost.


Building on these considerations, in this paper, we investigate near-field XL-MIMO systems employing sparse UPAs, which is a more practical and scalable architecture than ULAs.
We derive closed-form expressions for the signal beam power under various conditions to analyze the distance-focusing property and grating lobe suppression behavior as a function of antenna spacing. Furthermore, for single-user XL-MIMO systems equipped with sparse UPAs, we propose an algorithm to accurately estimate the effective degrees of freedom (EDoF) with low computational complexity.
The main contributions of this paper are summarized as follows:
\begin{itemize}
	\item We investigate a downlink near-field XL-MIMO system with sparse UPAs. We derive a closed-form expression for the power distribution of the signal beam along radial direction near the focused point $\mathbf{r}_0$. Then a closed-form expression for the length of the main lobe is derived. It shows the relationship between the distance-focusing property and the antenna spacing.
	\item We derive closed-form expressions for the signal beam power at a fixed propagation distance, explicitly characterizing the intensity of the grating lobes introduced by the sparse UPA. Building on this result, we further obtain a closed-form expression for the grating lobe suppression ratio and analyze its behavior, demonstrating that the suppression is enhanced when the focal distance decreases or the antenna spacing increases.
	\item We introduce a critical constraint on the system parameters under which the EDoF of XL-MIMO systems with sparse UPAs can be precisely estimated. Based on that, we propose an algorithm to obtain a closed-form expression for EDoF, which can estimate EDoF with high accuracy and low computational complexity.
\end{itemize}

The remainder of this paper is organized as follows:
Section~\ref{SA_Section2} introduces a downlink near-field XL-MIMO system with sparse UPAs.
Section~\ref{Sec_focusing} studies the distance-focusing property.
Section~\ref{Sec_suppression} investigates the grating lobe suppression behavior.
Section~\ref{SA_Section4} proposes an algorithm to obtain a closed-form expression for EDoF.
Section~\ref{SA_Section5} provides numerical results to verified the main results of this paper.
Section~\ref{SA_Section6} gives a brief conclusion of this paper.

\emph{Notations:}
In this paper,
scalars, vectors and matrices are respectively denoted by lower case letters, bold lower case letters and bold upper case letters.
The matrix inverse, conjugate-transpose, transpose and conjugate operations are respectively denoted by the superscripts ${\left(  \cdot  \right)^ {-1} }$, ${\left(  \cdot  \right)^H}$, ${\left(  \cdot  \right)^T}$ and ${\left(  \cdot  \right)^{*}}$.
We use ${\rm tr}\left(  \cdot  \right)$, $\left\|  \cdot  \right\|$ and ${\rm E}\left\{  \cdot  \right\}$ to denote trace, Euclidean 2-norm and the expectation operations, respectively.



\begin{figure*}[b]
	\hrulefill
	\setcounter{equation}{3}
\begin{align}
   & r_{m,n}=\sqrt{(r\sin \theta \cos \phi -dm)^2+(r\sin \theta \sin \phi -dn)^2+(r\cos \theta )^2}  \label{diatance_rmn}\\
   & \approx r-dm\sin \theta \cos \phi -dn\sin \theta \sin \phi +\frac{d^2m^2}{2r}\left( \cos ^2\theta +\sin ^2\theta \sin ^2\phi \right) +\frac{d^2n^2}{2r}\left( \cos ^2\theta +\sin ^2\theta \cos ^2\phi \right) -\frac{d^2mn\sin ^2\theta \cos \phi \sin \phi}{r} \label{diatance_rmn_taylor_second}  \\
   & \approx r-dm\sin \theta \cos \phi -dn\sin \theta \sin \phi +\frac{d^2m^2}{2r}\left( \cos ^2\theta +\sin ^2\theta \sin ^2\phi \right) +\frac{d^2n^2}{2r}\left( \cos ^2\theta +\sin ^2\theta \cos ^2\phi \right)  \label{diatance_rmn_approx}
\end{align}
	\setcounter{equation}{0}
\end{figure*}

\section{System Model}\label{SA_Section2}

Consider a downlink near-field XL-MIMO communication system as depicted in Fig. \ref{SA_p2}, where a sparse UPA is deployed at the base station (BS) with the size of $M\times N$ to transmit signals to users. We assume that no obstacles exist between the UPA and users, and thus only LoS path is considered.


Assuming that the XL-MIMO array transmits data symbol $s$ to the user at position $\mathbf{r}$ with the same transmit power of $\frac{P}{MN}$ for each antenna,
the arrived signal $f$ at position $\mathbf{r}$ can be expressed as
\begin{equation}\label{Sparse_eq1} 
	f=  \sqrt{\frac{P}{MN}} \sum_{m=-\frac{M-1}{2}}^{\frac{M-1}{2}}{\sum_{n=-\frac{N-1}{2}}^{\frac{N-1}{2}}{w^*_{m,n}h_{m,n}s}}  , 
\end{equation}
where the index $\left( m,n \right) $ represents the relative position for antennas in the UPA, and $\left( 0,0 \right) $ stands for the antenna at the center, as shown in Fig. \ref{SA_p2}.
Scalars $w_{m,n}$ and $h_{m,n}$ are respectively the beamforming coefficient and the channel coefficient for the antenna $\left( m,n \right) $.

To characterize the spherical wavefront in near-field, we adopt 
the Green's Function-based channel model, which is widely used in near-field XL-MIMO systems \cite{2019Waves}. 
Then, the channel coefficient $h_{m,n}$ is given by 
\begin{equation}\label{Sparse_eq2} 
	h_{m,n}=\frac{1}{4\pi r_{m,n}}e^{-j\frac{2\pi}{\lambda}r_{m,n}}\approx \frac{1}{4\pi r}e^{-j\frac{2\pi}{\lambda}r_{m,n}},
\end{equation}
where $r_{m,n}$ represents the distance between $\mathbf{r}$ and the antenna $\left( m,n \right) $.
The approximation in \eqref{Sparse_eq2} holds due to the fact that the power variations over arrays are negligible in radiative near field compared to the phase variations \cite{Towards6G}.
Additionally, considering maximal ratio (MR) beamforming, 
when the UPA focuses on the position $\mathbf{r}$, the normalized beamforming coefficient $w_{m,n}$
is given by 
\begin{equation}
	w_{m,n}=\frac{1}{\sqrt{MN}}e^{-j\frac{2\pi}{\lambda}r_{m,n}}.
\end{equation}

The coordinate of position $\mathbf{r}$ can be denoted as 
$(r\sin \theta \cos \phi , r\sin \theta \sin \phi , r\cos \theta )$,
where $r$ represents the distance between $\mathbf{r}$ and the antenna $\left( 0,0 \right) $. Angles $\theta$ and $\phi$ are respectively the elevation angle from the $Z_+$ axis and the azimuth angle from the $X_+$ axis.
Then, 
distance $r_{m,n}$ can be expressed as
\eqref{diatance_rmn} at the bottom of this page, 
where $d$ is antenna spacing.
$m$ and $n$ are defined over the ranges $m=-\frac{M-1}{2},\dots ,\frac{M-1}{2}$
and $n=-\frac{N-1}{2},\dots ,\frac{N-1}{2}$.
However, the square root expression poses analytical challenges for further performance analysis. To address this, we employ a second-order Taylor approximation, $\sqrt{1+u} \approx 1 + \frac{u}{2} - \frac{u^2}{8}$, retaining terms up to second order, which leads to equation \eqref{diatance_rmn_taylor_second} at the bottom of this page. Furthermore, we neglect the last cross term due to its negligible yet analytically intractable contribution, thereby obtaining the simplified form given in \eqref{diatance_rmn_approx} at the bottom of this page.
\addtocounter{equation}{3}

\begin{figure*}[b]
	\hrulefill
	\setcounter{equation}{10}
	\begin{equation}\label{SA_T2eq2} 
	\rho _{\mathrm{d}}=\begin{cases}
		\frac{1}{\left( b_M \right) ^2}\frac{1}{\left( b_N \right) ^2}\left( C^2\left( b_M \right) +S^2\left( b_M \right) \right) \left( C^2\left( b_N \right) +S^2\left( b_N \right) \right) &, \;\mu >0\\
		1&, \;\mu =0\\
	\end{cases}
    \end{equation}
	\setcounter{equation}{6}
\end{figure*}

\section{Distance-Focusing Property}\label{Sec_focusing}

In contrast to far-field scenarios, XL-MIMO systems with sparse UPAs exhibit a distance-focusing property in the near-field region, which allows users to be distinguished based on their radial distances. To examine this characteristic, we analyze in this section the power distribution of the signal beam along the radial direction. Although sparse arrays introduce grating lobes, our study in this section focuses solely on the main lobe. It is worth noting that the findings also apply to grating lobes, as they result from the periodic behavior of the phase.

Assume that the sparse UPA focuses on the position 
$
\mathbf{r}_0(r_0\sin \theta _0\cos \phi _0,r_0\sin \theta _0\sin \phi _0,r_0\cos \theta _0)
$, 
which acquires the normalized beamforming coefficient $w_{m,n}$ to be set as
\begin{equation}\label{w0_expression}
	w_{m,n}=w_{m,n}^{0}=\frac{1}{\sqrt{MN}}e^{-j\frac{2\pi}{\lambda}r_{m,n}^{0}}
\end{equation}
with  distance $r_{m,n}^{0}=r_{m,n}\left( r_0,\theta _0,\phi _0 \right) $.
To examine the length of the main lobe,  
we consider the signal \( f_{\mathrm{d}} \) arriving at the position  
$
\mathbf{r}_{\mathrm{d}} (r \sin \theta_0 \cos \phi_0, r \sin \theta_0 \sin \phi_0, r \cos \theta_0)
$,
where \( r = r_0 + r_e \).  
Here, \( \mathbf{r}_{\mathrm{d}} \) denotes any point sharing the same angular coordinates \( (\theta_0, \phi_0) \), and \( r_e \) represents the radial displacement relative to the distance \( r_0 \) of the focused position \( \mathbf{r}_0 \).
The channel coefficient for \( \mathbf{r}_{\mathrm{d}} \) can be expressed as
\begin{equation}\label{hd_expression}
	h_{m,n} = h_{m,n}^{\mathrm{d}}\approx \frac{1}{4\pi r}e^{-j\frac{2\pi}{\lambda}r_{m,n}^{\mathrm{d}}}
\end{equation}
with distance $r_{m,n}^{\mathrm{d}}=r_{m,n}\left( r,\theta _0,\phi _0 \right) $.
Consequently, the arrived signal $f_{\mathrm{d}}$ at position $ \mathbf{r}_{\mathrm{d}} $ can be expressed as
\begin{equation}\label{fd_expression}
	f_{\mathrm{d}}=\sqrt{\frac{P}{MN}}\sum_{m=-\frac{M-1}{2}}^{\frac{M-1}{2}}{\sum_{n=-\frac{N-1}{2}}^{\frac{N-1}{2}}{\left( w_{m,n}^{0} \right) ^{\ast}h_{m,n}^{\mathrm{d}}s}}.
\end{equation}

Then, a closed-form expression is derived for the power of $f_{\mathrm{d}}$ in the following theorem.

\begin{theorem}\label{Theorem_distance}
	When the XL-MIMO array focuses on position $
	\mathbf{r}_0(r_0\sin \theta _0\cos \phi _0,r_0\sin \theta _0\sin \phi _0,r_0\cos \theta _0)
	$, 
	the power $P_{\mathrm{d}}$ of the signal $f_{\mathrm{d}}$ arrived at $
	\mathbf{r}_{\mathrm{d}} (r \sin \theta_0 \cos \phi_0, r \sin \theta_0 \sin \phi_0, r \cos \theta_0)	$, which shares the same angular coordinates with $\mathbf{r}_0$,
	can be approximated as 
	\begin{equation}\label{Theorem_distance_Pd_approx}
		P_{\mathrm{d}}\approx \frac{P}{\left( 4\pi \left( r_0+r_e \right) \right) ^2}\rho _{\mathrm{d}},
	\end{equation}
	where $\rho _{\mathrm{d}}$ is expressed as \eqref{SA_T2eq2} at the bottom of this page,
with parameters given by  \addtocounter{equation}{1}
	\begin{equation*}
		b_M=\frac{M-1}{2}\tau _x\mu ,
	\end{equation*}
	\begin{equation*}
		b_N=\frac{N-1}{2}\tau _y\mu ,
	\end{equation*}
	\begin{equation*}
		\tau _x=\sqrt{\left( \cos ^2\theta _0+\sin ^2\theta _0\sin ^2\phi _0 \right)},
	\end{equation*}
	\begin{equation*}
	    \tau _y=\sqrt{\left( \cos ^2\theta _0+\sin ^2\theta _0\cos ^2\phi _0 \right)},
	\end{equation*}
	\begin{equation}\label{Theorem1_parameters} 
		\mu =d\sqrt{\frac{2}{\lambda}\left| \frac{r_e}{r_0\left( r_0+r_e \right)} \right|}\,\,.
	\end{equation}
	Functions $C(\cdot )$ and $S(\cdot )$ are Fresnel integrals, respectively expressed as
	\begin{equation*}
		C(x)=\int_0^x{\cos \left( \frac{\pi}{2}t^2 \right) dt},
	\end{equation*}
	\begin{equation}\label{Theorem1_CandS} 
	    S(x)=\int_0^x{\sin \left( \frac{\pi}{2}t^2 \right) dt}.
	\end{equation}
\end{theorem}

\begin{IEEEproof}
	Please refer to Appendix \ref{app_distance}.
\end{IEEEproof}

Theorem \ref{Theorem_distance} describes the power distribution of the signal beam at arbitrary positions along the radial direction that share the same angular coordinates as the focal point $\mathbf{r}_0$.
The factor $\frac{P}{\left( 4\pi \left( r_0+r_e \right) \right) ^2}$ in \eqref{Theorem_distance_Pd_approx} depicts the power variation induced by the path loss, 
while the factor $\rho_{\mathrm{d}}$ characterizes the power variation caused by the signal phase variation across the UPA.
It is noted from \eqref{SA_T2eq2} that $\rho_{\mathrm{d}}$ has a complex expression, and thus useful information regarding the length of the main lobe cannot be directly obtained.
To address that, 
we investigate the characteristic of $\rho _{\mathrm{d}}$ in the following corollary.

\begin{theorem}\label{SA_C1}
	For small values of $\mu$, the function $\rho_{\mathrm{d}}(\mu)$ defined in Eq. \eqref{SA_T2eq2} decreases monotonically as $\mu$ increases. For large $\mu$, the function exhibits bounded oscillations within a finite interval.
\end{theorem}

\begin{IEEEproof}
	Please refer to Appendix \ref{app_rho_d}.
\end{IEEEproof}

Theorem \ref{SA_C1} indicates that $\rho_{\mathrm{d}}(\mu)$ attains its peak near $\mu = 0^+$, with the maximum value occurring at $\mu = 0$. As described by \eqref{Theorem_distance_Pd_approx}, this peak characterizes the shape of the main lobe of the arrived signal along the radial direction centered at $\mathbf{r}_0$.
Therefore,
the width of the peak can be represented by the local minimum point $ \mu_{\min}$  closest to $\mu = 0$,
which can be obtained by Algorithm \ref{SA_Algo1}.

\begin{remark}
Moreover, according to \eqref{app_derivative_pho_d} and given that \(b_M\) and \(b_N\) are proportional to \(\mu\) as indicated in \eqref{Theorem1_parameters}, the parameter \(\mu_{\min}\) is approximately inversely proportional to the number of antennas. This relationship can be expressed as  
\begin{equation}  
	\mu_{\min} \approx \frac{c}{\max\left(M\tau_x, N\tau_y\right)},  
\end{equation}  
where \(c =  3.9\) is a constant.
\end{remark}

From the width of the peak $ \mu_{\min}$ and \eqref{Theorem1_parameters}, 
the extent of the main lobe centered at $\mathbf{r}_0$ along the $r$-axis can be determined, defined by its endpoints $r_{e,+}$ and $\bar r_{e,-}$, which are given by
\begin{equation*}
	r_{e,+}=\frac{\lambda \mu _{\min}^{2}r_{0}^{2}}{2d^2-\lambda \mu _{\min}^{2}r_0},
\end{equation*}
\begin{equation}\label{SA_section3_B_eq3}
	r_{e,-}=\frac{-\lambda \mu _{\min}^{2}r_{0}^{2}}{2d^2+\lambda \mu _{\min}^{2}r_0}.
\end{equation}
Equation \eqref{SA_section3_B_eq3} provides closed-form expressions for the endpoints of the main lobe in near-field XL-MIMO UPAs.
It should be aware that, 
to ensure the main lobe of near-field XL-MIMO UPAs concentrates at 
the focused point $\mathbf{r}_0$ along the $r$-axis, the system parameters need to satisfy $r_{e,+}>0$ and $r_{e,-}<0$.
Meanwhile, it is also worth noting that
when the main lobe of an XL-MIMO UPA is concentrated at the focal point $\mathbf{r}_0$ along the $r$-axis, the UPA gains the ability to distinguish users based on distance, with the extent of this ability inversely proportional to the length of the main lobe along the $r$-axis.
Thus, we have the following corollary.

\begin{corollary}\label{SA_C2}
	When the XL-MIMO array has the ability to distinguish users based on distance, the system parameters need to satisfy
	\begin{equation}\label{SA_C2_eq1}
		2d^2-\lambda \mu _{\min}^{2}r_0>0.
	\end{equation}
	Furthermore, the extent of this ability is inversely proportional to the length of the main lobe along the 
	$r$-axis near the focused point $\mathbf{r}_0$, which is expressed as
	\begin{equation}\label{SA_C2_eq2}
	 	r_{\mathrm{length}}=r_{e+}-r_{e-}=\frac{4\lambda \mu _{\min}^{2}r_{0}^{2}d^2}{(2d^2-\lambda \mu _{\min}^{2}r_0)(2d^2+\lambda \mu _{\min}^{2}r_0)}.
	\end{equation}
	
\end{corollary}

\begin{IEEEproof}
	The results can be obtained by \eqref{SA_section3_B_eq3} and the analysis below it.
\end{IEEEproof}

\begin{figure*}[b]
	\hrulefill
	\setcounter{equation}{24}
	\begin{equation}\label{Theorem_angle_rho_a_approx} 
		\rho _{\mathrm{a}}=\frac{1}{4a_x4a_y}\left( \left( C\left( u_{1,x} \right) +C\left( u_{2,x} \right) \right) ^2+\left( S\left( u_{1,x} \right) +S\left( u_{2,x} \right) \right) ^2 \right) \left( \left( C\left( u_{1,y} \right) +C\left( u_{2,y} \right) \right) ^2+\left( S\left( u_{1,y} \right) +S\left( u_{2,y} \right) \right) ^2 \right) 
	\end{equation}
	\begin{equation}\label{Theorem_angle_ux} 
		u_{1,x}=\sqrt{\left| a_x \right|}\left( M-1 \right) +\frac{b_x}{\sqrt{\left| a_x \right|}},   u_{2,x}=\sqrt{\left| a_x \right|}\left( M-1 \right) -\frac{b_x}{\sqrt{\left| a_x \right|}}
	\end{equation}
	\begin{equation}\label{Theorem_angle_abx} 
		a_x=\frac{d^2}{2r_0\lambda}\left( \cos ^2\theta _0-\cos ^2\theta +\sin ^2\theta _0\sin ^2\phi _0-\sin ^2\theta \sin ^2\phi \right) ,   b_x=\frac{d}{\lambda}\left( \sin \theta \cos \phi -\sin \theta _0\cos \phi _0 \right) 
	\end{equation}
	\begin{equation}\label{Theorem_angle_uy} 
		u_{1,y}=\sqrt{\left| a_y \right|}\left( N-1 \right) +\frac{b_y}{\sqrt{\left| a_y \right|}},   u_{2,y}=\sqrt{\left| a_y \right|}\left( N-1 \right) -\frac{b_y}{\sqrt{\left| a_y \right|}}
	\end{equation}
	\begin{equation}\label{Theorem_angle_aby} 
		a_y=\frac{d^2}{2r_0\lambda}\left( \cos ^2\theta _0-\cos ^2\theta +\sin ^2\theta _0\cos ^2\phi _0-\sin ^2\theta \cos ^2\phi \right) ,   b_y=\frac{d}{\lambda}\left( \sin \theta \sin \phi -\sin \theta _0\sin \phi _0 \right)  
	\end{equation}
	\hrulefill
	\begin{equation}\label{Pa_theta_sim}
		P_{\mathrm{a},\theta}=\frac{P}{M^2\left( 4\pi r_0 \right) ^2}\left| \sum_{m=-\frac{M-1}{2}}^{\frac{M-1}{2}}{e^{j2\pi \left( m\frac{d}{\lambda}\left( \sin \theta _0-\sin \theta \right) +m^2\frac{d^2}{2r_0\lambda}\left( \cos ^2\theta -\cos ^2\theta _0 \right) \right)}} \right|^2
	\end{equation}
	
	\setcounter{equation}{17}
\end{figure*}

Corollary \ref{SA_C2} provides many interesting insights for XL-MIMO systems with UPAs, which are discussed in the following paragraphs.

\begin{algorithm}[t]	
	\caption{Finding the Local Minimum Point}
	\begin{algorithmic}[1]\label{SA_Algo1}
		\STATE \textbf{Input}: Initial value $\mu = 0.0001$, step size $\Delta \mu = 0.01$, number of antennas $M,N$
		\STATE \textbf{Input}: Function $\rho_{\mathrm{d}}(\mu, M, N)$
		\STATE $\rho_{\text{prev}} = \rho_{\mathrm{d}}(\mu, M, N)$;
		\WHILE{true}
		\STATE $\mu_{\text{new}} = \mu + \Delta \mu$;
		\STATE $\rho_{\text{new}} = \rho_{\mathrm{d}}(\mu_{\text{new}}, M, N)$;
		
		\IF{$\rho_{\text{new}} < \rho_{\text{prev}}$}
		\STATE $\mu = \mu_{\text{new}}$;
		\STATE $\rho_{\text{prev}} = \rho_{\text{new}}$;
		\ELSE
		\STATE \textbf{Output} $\mu_{\min} = \mu$ \text{ as a local minimum point}
		\STATE \textbf{Stop}
		\ENDIF
		\ENDWHILE
		\STATE \textbf{Return}: Local minimum point $\mu_{\min}$
	\end{algorithmic}
\end{algorithm}

\subsubsection{For antenna spacing} Focusing the antenna spacing $d$, we can transform the constraint \eqref{SA_C2_eq1} as
\begin{equation}\label{SA_section3_B_eq3_1}
	 d>\mu _{\min}\sqrt{\frac{\lambda r_0}{2}}=\frac{c}{\max \left( M\tau _x,N\tau _y \right)}\sqrt{\frac{\lambda r_0}{2}}.
\end{equation}
It indicates that, 
given the number of the antennas $M,N$ and the focal distance $r_0$, 
the concentration of the main lobe  along the  $r$-axis near the focused point will not be observed in XL-MIMO systems with sparse UPAs until $d$ exceeds a certain threshold given by \eqref{SA_section3_B_eq3_1}. 

This can be explained by \eqref{app_Pd_expansion} in Appendix~\ref{app_distance}.
A small value of $d$ results in a correspondingly small variation of the signal phase with respect to distance $r$. 
As a result, the phase variation induced by changes in $r$ causes minimal fluctuations in $\rho _{\mathrm{d}}$. 
This leads to the dominance of the path loss factor $\frac{P}{\left( 4\pi \left( r_0+r_e \right) \right) ^2}$ in $P_{\mathrm{d}}$ of \eqref{Theorem_distance_Pd_approx}, preventing the signal energy from being concentrated at the focused point $\mathbf{r}_0$ along the $r$-axis.

On the other hand, when the antenna spacing $d$ exceeds the constraint \eqref{SA_section3_B_eq3_1},
it is readily observed from \eqref{SA_C2_eq2} that the length of the main lobe decreases as $d$ increases, thereby the distance focusing ability being improved.

\subsubsection{For number of antennas}
Focusing the number of the antennas $N$ in the UPA, 
we transform the constraint \eqref{SA_C2_eq1} to
\begin{equation}\label{SA_section3_B_eq3_2}
	\mu _{\min}<d\sqrt{\frac{2}{\lambda r_0}}\Rightarrow \max \left( M\tau _x,N\tau _y \right) >\frac{c}{d}\sqrt{\frac{\lambda r_0}{2}}.
\end{equation}
Substituting $d = \lambda / 2$ into \eqref{SA_section3_B_eq3_2}, we have 
\begin{equation}\label{SA_section3_B_eq3_3}
	\max \left( M\tau _x,N\tau _y \right) >c\sqrt{\frac{2r_0}{\lambda}}.
\end{equation}
A valuable insight for the conventional collected UPA can be obtained
that, 
to ensure XL-MIMO systems with collected UPAs has the ability to distinguish users based on distance,
an excessive $M$ or $N$ is required, especially when high-frequency systems are considered.

In contrast, when sparse UPAs are considered, the required number of antennas can be rapidly reduced, bringing a promising reduction in hardware cost and energy consumption of XL-MIMO systems.

\subsubsection{For distance}
Focusing the distance $L$ between the UPA and the focused point, we transform the constraint \eqref{SA_C2_eq1} to
\begin{equation}\label{SA_section3_B_eq3_4}
	r_0<\frac{2d^2}{\lambda \mu _{\min}^{2}}=\frac{2d^2\left( \max \left( M\tau _x,N\tau _y \right) \right) ^2}{c^2\lambda}\approx \frac{r_{\mathrm{F}}}{c^2}\triangleq r_{\mathrm{RRD}},
\end{equation}
where $
r_{\mathrm{F}}=\frac{2D^2}{\lambda}\approx \frac{2d^2\left( \max \left( M\tau _x,N\tau _y \right) \right) ^2}{\lambda}
$ is the Fraunhofer (Rayleigh) distance \cite{7942128}, and $D$ is the effective aperture of the UPA.
In constraint \eqref{SA_section3_B_eq3_4}, we give the definition of $r_{\mathrm{RRD}}$, referred to as the radial resolution distance (RRD),
which represents the distance range within which XL-MIMO systems with UPAs can distinguish users based on distance, and outside of this range, they cannot.
It is observed that $r_{\mathrm{RRD}}$ is much smaller than the Fraunhofer distance $r_{\mathrm{F}}$.
Based on definition of $r_{\mathrm{RRD}}$, it may serve as a boundary instead of $r_{\mathrm{F}}$ to differentiate between the near-field and far-field regions in certain studies, such as near-field and far-field codebook design.

Furthermore, 
it is noted that the RRD is directly related to the square of the antenna spacing $d$, 
meaning the XL-MIMO systems with sparse UPAs are likely to have an outstanding ability of distinguishing users based on distance.

\section{Grating Lobe Suppression Behavior}\label{Sec_suppression}

In the far-field case, the grating lobes inherent to sparse arrays exhibit power levels comparable to that of the main lobe, resulting in substantial deterioration of system performance.
By contrast, in the near-field case, the intensity of grating lobes is naturally suppressed by using sparse arrays, which will be further investigated in this subsection.

Assume that the sparse UPA focuses on the position $\mathbf{r}_0$ with the normalized beamforming coefficient $w^0_{m,n}$ given by \eqref{w0_expression}.
To study the powers of the main lobe and grating lobes, 
we consider the signal $f_{\mathrm{a}}$ arriving at the position
$
\mathbf{r}_{\mathrm{a}}(r_0\sin \theta \cos \phi ,r_0\sin \theta \sin \phi ,r_0\cos \theta)
$.  
Here, \( \mathbf{r}_{\mathrm{a}} \) denotes any point sharing the same distance $r_0$ with the focused position \( \mathbf{r}_0 \), but having different 
angular coordinates \( (\theta, \phi) \).
The channel coefficient for \( \mathbf{r}_{\mathrm{a}} \) can be expressed as
\begin{equation}\label{ha_expression}
	h_{m,n}=h_{m,n}^{\mathrm{a}}\approx \frac{1}{4\pi r}e^{-j\frac{2\pi}{\lambda}r_{m,n}^{\mathrm{a}}}
\end{equation}
with distance $r_{m,n}^{\mathrm{a}}=r_{m,n}\left( r_0,\theta ,\phi  \right) $.
Consequently, the arrived signal $f_{\mathrm{d}}$ at position $ \mathbf{r}_{\mathrm{d}} $ can be expressed as
\begin{equation}\label{fa_expression}
	f_{\mathrm{a}}=\sqrt{\frac{P}{MN}}\sum_{m=-\frac{M-1}{2}}^{\frac{M-1}{2}}{\sum_{n=-\frac{N-1}{2}}^{\frac{N-1}{2}}{\left( w_{m,n}^{0} \right) ^{\ast}h_{m,n}^{\mathrm{a}}s}}.
\end{equation}

Firstly, we derive a closed-form expression for the power of the main lobe of $f_{\mathrm{a}}$ in the following theorem.

\begin{theorem}\label{Theorem_angle}
    When the XL-MIMO array focuses on position $
	\mathbf{r}_0(r_0\sin \theta _0\cos \phi _0,r_0\sin \theta _0\sin \phi _0,r_0\cos \theta _0)
	$, 
	the power $P_{\mathrm{a}}$ of the main lobe of  the signal $f_{\mathrm{a}}$ arrived at $
	\mathbf{r}_{\mathrm{a}} (r_0 \sin \theta \cos \phi, r \sin \theta \sin \phi, r \cos \theta)	$, which shares the same distance coordinate with $\mathbf{r}_0$,
	can be approximated as 
	\begin{equation}\label{Theorem_angle_Pa_approx}
		P_{\mathrm{a}}\approx \frac{P}{M^2N^2\left( 4\pi r_0 \right) ^2}\rho _{\mathrm{a}},
	\end{equation}
	where $\rho _{\mathrm{a}}$ is expressed as \eqref{Theorem_angle_rho_a_approx} at the bottom of the previous page, with the parameters given by \eqref{Theorem_angle_ux} - \eqref{Theorem_angle_aby}.  \addtocounter{equation}{5}
\end{theorem}

\begin{IEEEproof}
	Please refer to Appendix \ref{app_angle}.
\end{IEEEproof}

Theorem \ref{Theorem_angle} illustrates the power distribution of the main lobe of the  signal $f_{\mathrm{a}}$ arrived at $\mathbf{r}_{\mathrm{a}}$, when the sparse UPA focuses on $\mathbf{r}_0$. 
The approximation in \eqref{Theorem_angle_Pa_approx} therefore requires that the angular coordinates \( (\theta, \phi) \) of \( \mathbf{r}_{\mathrm{a}} \) be close to those of \( \mathbf{r}_0 \), i.e., \( (\theta_0, \phi_0) \).  

To analyze the power distribution of grating lobes, we restrict our attention to signals arrived in the \(\mathrm{XoZ}\) plane to simplify the derivation. This corresponds to the case where \( \phi = \phi_0 = 0 \), and the exact expression for $P_{\mathrm{a}}$ reduces from \eqref{app_Pa_expansion} to the simplified form $P_{\mathrm{a},\theta}$ given in \eqref{Pa_theta_sim}.\addtocounter{equation}{1}
Based on this formulation, we first examine the angular location of the grating lobes, as presented in the following corollary.

\begin{corollary}\label{Corollary_grating_lobe_location}
   When the azimuth angle is set to \( \phi = \phi_0 = 0 \), the angular locations of the main lobe and grating lobes are given by
   \begin{equation}\label{Corollary_theta_k}
   	\theta _{k}^{\prime} = \arcsin \left( \sin \theta_0 + \frac{k\lambda}{d} \right),
   \end{equation}
   where \( k \) is an integer in the interval
   \begin{equation}\label{Corollary_k_range}
   	   k \in \left[ \, \left\lceil \left( -1 - \sin \theta_0 \right) \frac{d}{\lambda} \right\rceil ,\; \left\lfloor \left( 1 - \sin \theta_0 \right) \frac{d}{\lambda} \right\rfloor \, \right].
   \end{equation}
   Here, \( \lceil \cdot \rceil \) and \( \lfloor \cdot \rfloor \) denote the ceiling and floor functions, respectively. The case \( k = 0 \) corresponds to the main lobe, while non-zero values of \( k \) correspond to grating lobes.
   
\end{corollary}

\begin{IEEEproof}
	Please refer to Appendix \ref{app_angle_location}.
\end{IEEEproof}

\begin{figure*}[t]
	\setcounter{equation}{32}
	\begin{equation}\label{Pa_theta_approx} 
		P_{\mathrm{a},\theta}=\frac{P}{M^2\left( 4\pi r_0 \right) ^2}\frac{1}{4a_{x,\theta}}\left( \left( C\left( u_{1,x,\theta} \right) +C\left( u_{2,x,\theta} \right) \right) ^2+\left( S\left( u_{1,x,\theta} \right) +S\left( u_{2,x,\theta} \right) \right) ^2 \right)  
	\end{equation}
	\hrulefill
	
	\setcounter{equation}{41}
	 \begin{equation}\label{derivative_eta_k} 
		\frac{d}{d\zeta}\eta _k=\frac{2}{\zeta ^2}\left[ -\frac{1}{\zeta}\left( C^2(\zeta )+S^2(\zeta ) \right) +C(\zeta )\cos \left( \frac{\pi \zeta ^2}{2} \right) +S(\zeta )\sin \left( \frac{\pi \zeta ^2}{2} \right) \right] 
	\end{equation}
	\hrulefill
	
	\setcounter{equation}{32}
\end{figure*}

Corollary \ref{Corollary_grating_lobe_location} points out the angular location of the grating lobes. 
Based on that, we can further derive the closed-form expression for the power of the grating lobes.

\begin{theorem}\label{Theorem_grating_lobe_intensity}
	When the azimuth angle is set to \( \phi = \phi_0 = 0 \), the power of the main lobe and grating lobes can be approximated as \eqref{Pa_theta_approx} at the top of this page,
	where parameters $u_{1,x,\theta}$, $u_{2,x,\theta}$ and $a_{x,\theta}$ are respectively given by \addtocounter{equation}{1}
	\begin{equation}
		u_{1,x,\theta}=\sqrt{\left| a_{x,\theta} \right|}\left( M-1 \right) +\frac{b_{x,\theta}}{\sqrt{\left| a_{x,\theta} \right|}},
	\end{equation}
	\begin{equation}
		u_{2,x,\theta}=\sqrt{\left| a_{x,\theta} \right|}\left( M-1 \right) -\frac{b_{x,\theta}}{\sqrt{\left| a_{x,\theta} \right|}},
	\end{equation}
	\begin{equation}
		a_{x,\theta}=\frac{d^2}{2r_0\lambda}\left( \cos ^2\theta _0-\cos ^2\theta \right),
	\end{equation}
	\begin{equation}
		b_{x,\theta}=\left\{ \left( \theta ,\frac{d}{\lambda}\left( \sin \theta -\sin \theta _{k}^{\prime} \right) \right) \, \middle| \,\theta \in I_k \right\} , 
	\end{equation}
	with the interval $I_k$ being
	\begin{equation}
		I_k=\begin{cases}
			\left[ -\frac{\pi}{2},\frac{\theta _{k}^{\prime}+\theta _{k+1}^{\prime}}{2} \right] , k=\lceil \left( -1-\sin \theta _0 \right) \frac{d}{\lambda} \rceil\\
			\left[ \frac{\theta _{k-1}^{\prime}+\theta _{k}^{\prime}}{2},\frac{\pi}{2} \right] , k=\lfloor \left( 1-\sin \theta _0 \right) \frac{d}{\lambda} \rfloor\\
			\left[ \frac{\theta _{k-1}^{\prime}+\theta _{k}^{\prime}}{2},\frac{\theta _{k}^{\prime}+\theta _{k+1}^{\prime}}{2} \right] , else\\
		\end{cases}.
	\end{equation}
	Here, the expression of $\theta _{k}^{\prime}$ and the range of integer $k$ are respectively given by \eqref{Corollary_theta_k} and \eqref{Corollary_k_range}.
\end{theorem}

\begin{IEEEproof}
	Please refer to Appendix \ref{app_angle_intensity}.
\end{IEEEproof}

Theorem \ref{Theorem_grating_lobe_intensity} derives the closed-form expression for each of the grating lobes.
Based on that, we can further investigate the phenomenon of the grating lobe suppression with sparse arrays in the near-field.

\begin{corollary}\label{Corollary_grating_lobe_suppression}
	When the azimuth angle is set to \( \phi = \phi_0 = 0 \), 
	the ratio of the peak power of the grating lobe at \( \theta _{k}^{\prime} \) to that of the main lobe at \( \theta _0 \) is given by  
	\begin{equation}\label{eta_k}
		\eta _k=\frac{P_{\mathrm{a},\theta}\left( \theta _{k}^{\prime} \right)}{P_{\mathrm{a},\theta}\left( \theta _0 \right)}
		= \frac{1}{\zeta ^2}\left( C^2\left( \zeta \right) +S^2\left( \zeta \right) \right) ,
	\end{equation}
	where $\zeta $ is defined as
	\begin{equation}\label{zeta_definition}
		\zeta =\left( M-1 \right) \sqrt{\frac{1}{r_0}\left| dk\sin \theta _0+\frac{1}{2}k^2\lambda \right|} .
	\end{equation}

\end{corollary}

\begin{IEEEproof}
	Substituting $\theta = \theta _{k}^{\prime}$ and $\theta = \theta _0$ into \eqref{Pa_theta_approx} respectively, we can easily obtain \eqref{eta_k} with
	\begin{equation}\label{zeta_definition_temp}
		\zeta =\left( M-1 \right) \sqrt{\frac{d^2}{2r_0\lambda}\left| \cos ^2\theta _0-\cos ^2\theta _{k}^{\prime} \right|} .
	\end{equation}
	Then substituting \eqref{Corollary_theta_k} into \eqref{zeta_definition_temp}, we arrive at \eqref{zeta_definition} and complete the proof. 
\end{IEEEproof}

Corollary \ref{Corollary_grating_lobe_suppression} quantifies the degree of power suppression for the grating lobes.  
This can be further elucidated by evaluating the derivative given in \eqref{derivative_eta_k} at the top of this page.   \addtocounter{equation}{1}
When \(\zeta\) is close to zero, the term involving \(-\frac{1}{\zeta}\) dominates the sign of the derivative, as the remaining terms remain bounded.  
Consequently, \(\eta_k\) decreases as \(\zeta\) increases from zero, indicating that the power of the grating lobes is suppressed.

Furthermore, from the expression of \(\zeta\) in \eqref{zeta_definition}, it can be seen that \(\zeta\) increases when the distance \(r_0\) decreases.  
This implies that grating lobes generated from sparse arrays can be effectively suppressed in the near field.  
In contrast, in the far-field case where \(r_0 \rightarrow \infty\), we have \(\zeta \rightarrow 0\) and \(\lim_{\zeta \rightarrow 0} \eta_k = 1\), indicating no grating lobe suppression, as the grating lobe and main lobe possess equal peak power.

Moreover, according to \eqref{zeta_definition}, in the near field with sparse arrays, increasing the antenna spacing \(d\) leads to the suppression of most grating lobes.  
Therefore, particular attention should be paid to the few grating lobes that retain relatively high power.  
As discussed earlier, the suppression ratio is maximized when \(\zeta \rightarrow 0\).
Thus, we arrive at the following corollary:

\begin{corollary}\label{Corollary_grating_lobe_index}
	When the azimuth angle is set to \( \phi = \phi_0 = 0 \), 
	the indices of the most powerful grating lobes are given by    
	\begin{equation}\label{index_k}
		k=\lfloor -2\frac{d}{\lambda}\sin \theta _0 \rfloor \,\,\mathrm{or}\,\, \lfloor -2\frac{d}{\lambda}\sin \theta _0 \rfloor +1 .
	\end{equation}
	
\end{corollary}

\begin{IEEEproof}
	The most powerful grating lobes are those with a suppression ratio $\eta_k$ approaching unity, which requires \(\zeta \rightarrow 0\).  
	From \eqref{zeta_definition}, it follows that
	\begin{equation}\label{index_k_temp}
		\zeta = 0 \quad \Longrightarrow \quad k = -2\frac{d}{\lambda}\sin \theta_0.
	\end{equation}
	Furthermore, it can be observed that
	\begin{equation}
		\left( -1 - \sin \theta_0 \right) \frac{d}{\lambda}
		\leqslant -2\frac{d}{\lambda}\sin \theta_0
		\leqslant 
		\left( 1 - \sin \theta_0 \right) \frac{d}{\lambda},
	\end{equation}
	indicating that the value in \eqref{index_k_temp} lies within the required range specified by \eqref{Corollary_k_range}.  
	Since the index $k$ must be an integer, we obtain \eqref{index_k}, which completes the proof.
\end{IEEEproof}

Corollary~\ref{Corollary_grating_lobe_index} identifies the indices of the most powerful grating lobes in near-field XL-MIMO systems employing UPAs.  
While increasing the antenna spacing can effectively suppress other grating lobes, those corresponding to the indices given in \eqref{index_k} remain largely unaffected.  
This underscores the importance of developing alternative methods to suppress these particular grating lobes, which is essential for advancing the application of sparse UPAs in future near-field communications.

Sections \ref{Sec_focusing} and \ref{Sec_suppression}  reveal two beneficial properties of using sparse UPAs in XL-MIMO systems: distance-focusing property and grating lobe suppression behavior, both of which are enhanced by increasing the antenna spacing \(d\). As a result, the area and intensity of the grating lobes can be substantially reduced, leading to significant mitigation of inter-user interference in multi-user scenarios. These findings support the practical feasibility of deploying sparse arrays in near-field XL-MIMO systems.

%
%

\section{EDoF Function}\label{SA_Section4}

A critical advantage of XL-MIMO is that 
it can largely improve the EDoF of communication systems
due to the spherical wavefront in near-field.
In this section, we consider a single-user XL-MIMO communication system, where the transmit array and receive array deploy sparse UPAs with the antenna spacing denoted as $d$ and $\bar d$, respectively.

\begin{figure}[t]
	\centering
	\includegraphics[scale=0.50]{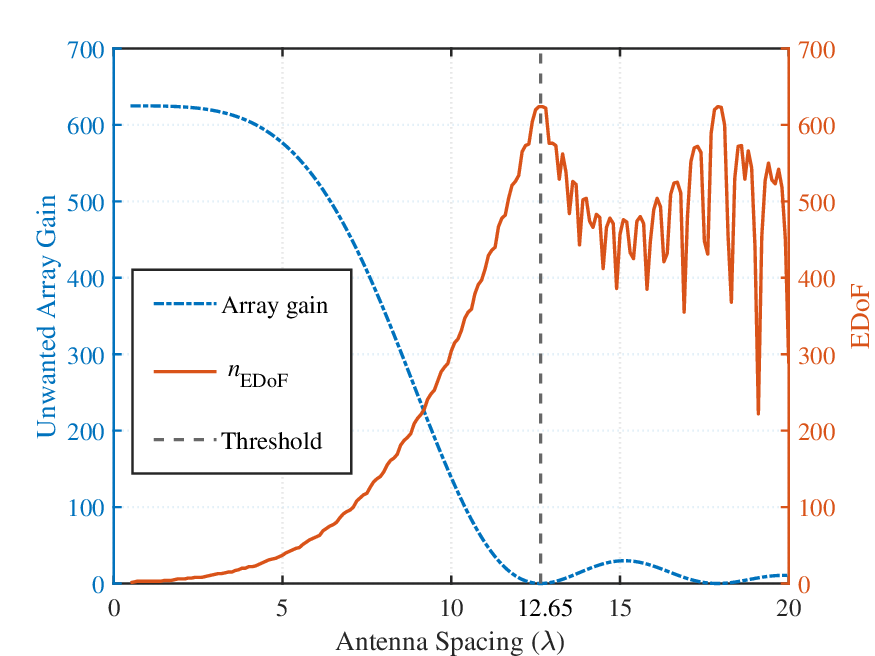}\\  
	\caption{Relationship between array gain and EDoF.}\label{SA_p4} 
\end{figure}

As demonstrated in our previous work \cite{chen2025}, the use of a sparse array configuration can lead to substantial improvement in EDoF as shown in Fig.~\ref{SA_p4}. 
The EDoF gradually reaches its maximum when the unwanted array gain decreases to zero. This condition is achieved by increasing the antenna spacing, which reduces the gain measured at the antenna nearest to the focused antenna in the receive array.

\subsection{Existing EDoF Estimation Methods}\label{sec4-A}

Since EDoF is significantly increased in XL-MIMO systems with sparse arrays, methods to estimate EDoF become crucial, which have drawn extensive research attention.

The direct solution to obtain EDoF is to count the number of significant singular values of $\mathbf{G}$, which is the channel matrix between the transmitter and receiver.
This process can be described by 
\begin{equation}\label{SA_section4_B_eq1} 
 n_{\mathrm{EDoF}}=\underset{n}{\text{argmin}}\left\{ f\left( n \right) =\sum_{i=1}^n{\mu _{i}^{2}} \; \Bigg| \; \frac{f\left( n \right)}{\sum_{i=1}^{n_{\mathrm{DoF}}}{\mu _{i}^{2}}}  \geq 99.9\%  \right\}   ,
\end{equation}
where $\mu _i$ represents the $i$-th largest singular value of $\mathbf{G}$.
It is noted that the direct solution does not give a closed-form expression for EDoF.
Besides, $\mu _i$ is obtained by the SVD operation, which has a high computational complexity.

One of the widely used methods to estimate the EDoF is given by \cite{2019Waves} 
\begin{equation}\label{SA_section4_B_eq2} 
  n_{\mathrm{EDoF1}}=\frac{A_{\mathrm{S}}A_{\mathrm{R}}}{\lambda ^2L^2},
\end{equation}
where $A_{\mathrm{S}}$ and $A_{\mathrm{R}}$ are the areas of the transmit and receive arrays, respectively.
This method provides a concise closed-form expression for EDoF, 
but its accuracy is limited.

Another method can estimate EDoF providing relatively better accuracy with the closed-form expression given by \cite{9650519} 
\begin{equation}\label{SA_section4_B_eq3} 
  n_{\mathrm{EDoF}2}=\frac{\mathrm{tr}^2\left( \mathbf{GG}^H \right)}{\left\| \mathbf{GG}^H \right\| _{\mathrm{F}}^{2}}=\frac{\left( \sum_i{\mu _{i}^{2}} \right) ^2}{\sum_i{\mu _{i}^{4}}}.
\end{equation}
However, this estimation method still has the drawback of high computational complexity.

\begin{table*}[b]
	\centering
	\caption{Coefficients of the EDoF fitting function.}
	\label{SA_table1}
	\begin{tabular}{c|cccccc}
		\toprule
		& $1$ & $\left(\frac{r}{\lambda}\right)^1$ & $\left(\frac{r}{\lambda}\right)^2$ & $\left(\frac{r}{\lambda}\right)^3$ & $\left(\frac{r}{\lambda}\right)^4$ & $\left(\frac{r}{\lambda}\right)^5$ \\
		\midrule
		$1$ & $p_{00} = 44.17$ & $p_{01} = -0.04529$ & $p_{02} = 1.963 \times 10^{-5}$ & $p_{03} = -3.933 \times 10^{-9}$ & $p_{04} = 3.645 \times 10^{-13}$ & $p_{05} = -1.265 \times 10^{-17}$ \\
		$\cos(\theta)$ & $p_{10} = 108.8$ & $p_{11} = -0.06321$ & $p_{12} = 1.438 \times 10^{-5}$ & $p_{13} = -1.422 \times 10^{-9}$ & $p_{14} = 5.241 \times 10^{-14}$ & $0$ \\
		$\cos^2(\theta)$ & $p_{20} = -18.19$ & $p_{21} = 0.01045$ & $p_{22} = -1.764 \times 10^{-6}$ & $p_{23} = 4.984 \times 10^{-11}$ & $0$ & $0$ \\
		$\cos^3(\theta)$ & $p_{30} = -21.37$ & $p_{31} = 0.002555$ & $p_{32} = 4.038 \times 10^{-7}$ & $0$ & $0$ & $0$ \\
		$\cos^4(\theta)$ & $p_{40} = 20.04$ & $p_{41} = -0.00338$ & $0$ & $0$ & $0$ & $0$ \\
		$\cos^5(\theta)$ & $p_{50} = -2.397$ & $0$ & $0$ & $0$ & $0$ & $0$ \\
		\bottomrule
	\end{tabular}
\end{table*}

\subsection{EDoF Function Fitting}

As is discussed in Subsection \ref{sec4-A}, the closed-form expressions in existing EDoF estimation methods are whether with limited accuracy or with high computational complexity.
To address that, we aim to propose a method in this subsection to obtain a closed-form expression,
which can estimate EDoF in XL-MIMO systems with high accuracy and low computational complexity.

As shown in Fig. \ref{SA_p4}, the EDoF increases steadily until reaching a peak at a specific antenna spacing threshold, beyond which it exhibits irregular fluctuations. This behavior implies that, within the range below this threshold, an accurate expression for the EDoF can be derived through function fitting in terms of the system parameters. This leads to the following theorem.

\begin{theorem}\label{Theorem_EDoF_constraint}
	For a single-user XL-MIMO system with an $N \times N$ transmit UPA, an accurate expression for the EDoF can be obtained by function fitting using system parameters, provided that the following constraint is satisfied:
	\begin{equation}\label{SA_section4_C_eq1}
		\frac{d \bar{d} N}{\lambda r} < 1,
	\end{equation}
	where \(d\) and \(\bar{d}\) denote the antenna spacing of the transmit and receive arrays, respectively.
\end{theorem}

\begin{IEEEproof}
	The antenna spacing threshold introduced in \cite[Theorem~1]{chen2025} for an \(N \times N\) transmit UPA is given by
	\[
	d_{\mathrm{threshold}} = \frac{\lambda r}{\bar{d} N}.
	\]
	The condition that the actual antenna spacing \(d\) remains below this threshold, i.e., \(d < d_{\mathrm{threshold}}\), is equivalent to
	\[
	\frac{d \bar{d} N}{\lambda r} < 1,
	\]
	which is exactly the constraint stated in \eqref{SA_section4_C_eq1}. Within this regime, the EDoF varies smoothly with the system parameters, enabling reliable function fitting.
\end{IEEEproof}

\begin{algorithm}[t]
	\caption{Obtain the EDoF Fitting Function}
	\begin{algorithmic}[1]\label{SA_Algo2}
		\STATE \textbf{INPUT:} \textbf{System parameters:} antenna spacing of the transmit array $d$, antenna spacing of the receive array $\bar{d}$, the number of transmit antennas $N$, system operating wavelength $\lambda$. 
		\textbf{Users' location information:} Range vector of users' angles $\bm{\theta }\in \left[ \theta _{\min}:\theta _{\mathrm{step}}:\theta _{\max} \right] $,
		Range vector of distance between the transmit array and users $
		\mathbf{r}\in \left[ r_{\min}:r_{\mathrm{step}}:r_{\max} \right] $
		
		\IF {$\frac{d\bar{d}{N}}{\lambda r_{\min}} < 1 $}
		
		\FOR{ $i=1:\mathrm{length}\left( \bm{\theta } \right) $}
		\FOR{ $j=1:\mathrm{length}\left( \mathbf{r} \right) $}
		
		\STATE Compute EDoF for the $\left( i,j \right) $ element of the EDoF matrix $\mathbf{H}_{\mathrm{EDoF}}$ using direct solution \eqref{SA_section4_B_eq1};
		
		\ENDFOR
		\ENDFOR
		
		\ELSE 
		\STATE \textbf{Print} ``Invalid Input'';
		
		\ENDIF
		
		\STATE \textbf{FITTING PROCESS:}
		\STATE Fit the relationship between EDoF matrix $\mathbf{H}_{\mathrm{EDoF}}$ and angle vector $\bm{\theta }$ and distance vector $\mathbf{r}$ by MATLAB tools, and obtain the EDoF fitting function $
		f_{\mathrm{EDoF}}\left( \theta ,r \right) $;
		
		\STATE \textbf{OUTPUT:} EDoF Fitting function $
		f_{\mathrm{EDoF}}\left( \theta ,r \right) $
	\end{algorithmic}
\end{algorithm}

In practice, it is important to obtain EDoF for receivers at different positions.
Therefore, we are interested in obtaining a closed-form expression for EDoF that relates the position of receivers. 
For the ease of analysis, we consider the case of $\phi = 0$, meaning that the  position of the receive array varies at the $\mathrm{XoZ}$ plane.
For the case that the system parameters satisfy the constraint \eqref{SA_section4_C_eq1},
we propose Algorithm \ref{SA_Algo2}
to obtain an EDoF fitting function $f_{\mathrm{EDoF}}\left( \theta ,r \right) $.
Then, the closed-form expression $f_{\mathrm{EDoF}}\left( \theta ,r \right) $ can estimate EDoF by the position information of the receiver array.

For example, we consider a system operating at $f=30$ GHz ($\lambda = 0.01$ m). 
The distance of interest between the transmit array and receive array lies in
$r \in \left[ 1000\lambda ,4000\lambda \right] $,
and the elevation angle satisfies
$\theta \in \left[ 0,\pi /2-\pi /30 \right] $.
The transmit UPA at the BS is with the size of $35 \times 35$ and antenna spacing $d = 10\lambda$.
The receive UPA is assumed to be in the XoZ plane, with the size of $9 \times 9$ and $\bar{d} = 2\lambda$.
Since the system parameters satisfy the constraint \eqref{SA_section4_C_eq1}, i.e., $\frac{d\bar{d}{N}}{\lambda r_{\min}} < 1 $, 
EDoF varies steadily with the receiver's position,
which means that an accurate expression for EDoF can be obtained by performing function fitting.
In this case, we can utilize Algorithm \ref{SA_Algo2} to obtain the EDoF fitting function, which is given by
\begin{equation}\label{SA_section5_eq1}
	f_{\mathrm{EDoF}}\left( \theta ,r \right) =\sum_{i=0}^5{\sum_{j=0}^5{p_{ij}\cos ^i\left( \theta \right) \left( \frac{r}{\lambda} \right) ^j}},
\end{equation}
where the coefficients are given by Table \ref{SA_table1}.

Algorithm \ref{SA_Algo2} provides an EDoF fitting function $f_{\mathrm{EDoF}}\left( \theta ,r \right) $ for receivers in the XoZ plane.
When the system parameters satisfy the constraint \eqref{SA_section4_C_eq1}, 
EDoF varies steadily with the system parameters,
making the function fitting yield excellent results.
Therefore, the EDoF fitting function $f_{\mathrm{EDoF}}\left( \theta ,r \right)$ is very accurate. 
Additionally, $f_{\mathrm{EDoF}}\left( \theta ,r \right)$ only depends on the position parameters $\theta$ and $r$, which indicates that, once the position information is obtained, the computation of EDoF entails minimal complexity.

\section{Numerical Results}\label{SA_Section5}

In this section, numerical results are provided to verify the main results of this paper and provide more insights.
We assume that the XL-MIMO system operates at 300 GHz, i.e., $\lambda = 0.001$ m.
A UPA is used for the XL-MIMO system with the size of $M \times N = 35 \times 35$.
The UPA is placed in the XoY plane, with its edges parallel to the coordinate axes and the center at $\left(0, 0, 0 \right)$.
The total transmit power is set as $P=10 \mathrm{dB}$.
Additionally, 
the default coordinates for the focused point $\mathbf{r}_0 $ in simulations are set as $r_0=5 \mathrm{m}$ and $\theta _0=\phi _0=0^\circ$, unless explicitly noted.

\begin{figure}[t]
	\centering
	\includegraphics[scale=0.50]{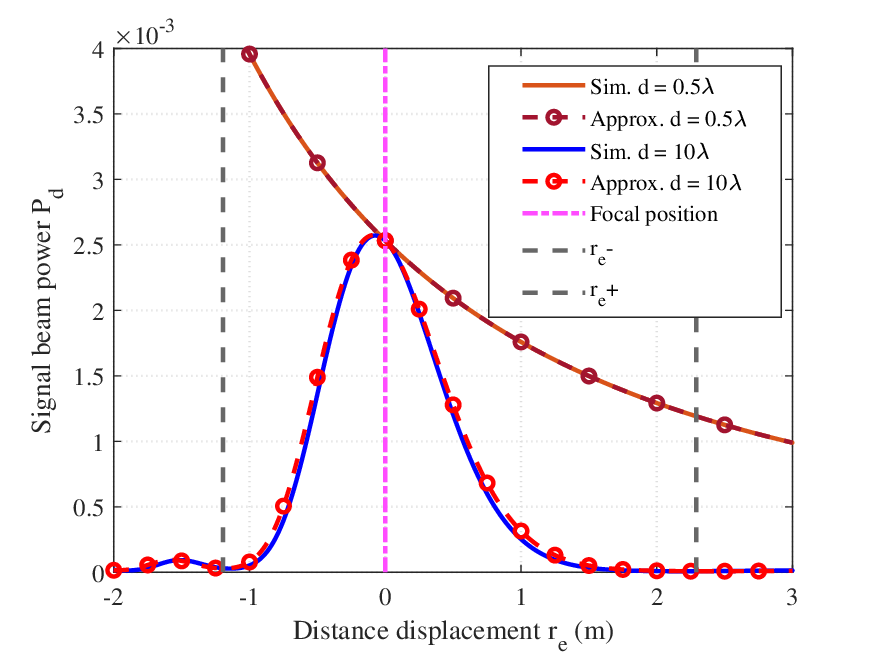}\\  
	\caption{Power distribution in the radial direction with $d = 0.5\lambda$ and $d = 10\lambda$.}\label{pic_Pd_re} 
\end{figure}

\begin{figure}[t]
	\centering
	\includegraphics[scale=0.34]{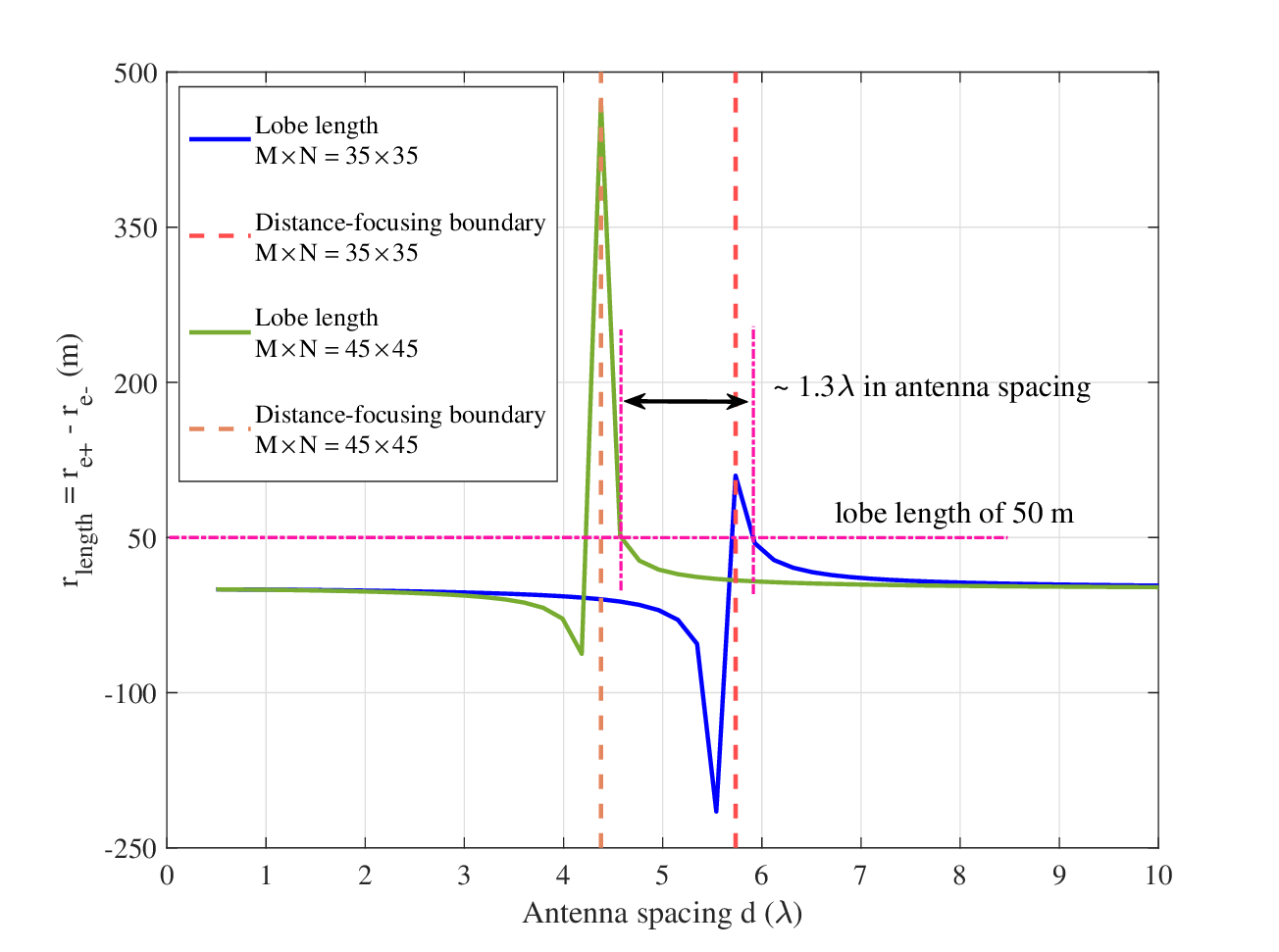}\\  
	\caption{Length of main lobe versus antenna spacing.}\label{pic_rlength_d} 
\end{figure}

Fig. \ref{pic_Pd_re} illustrates the power distribution of the arrived signal along radial direction near the focused point $\mathbf{r}_0$ with $d = 0.5\lambda$ and $d = 10\lambda$. 
The solid lines are based on the exact value of the power of the arrived signal, 
while the dashed lines are based on the approximation given by $P_{\mathrm{d}}$ in Theorem~\ref{Theorem_distance}.
It can be observed that the dashed lines well match the solid lines, indicating the correctness of Theorem~\ref{Theorem_distance}.
Two red vertical lines are drawn at $x= r_{e,-}$ and $x= r_{e,+}$ based on \eqref{SA_section3_B_eq3}, which exhibit the length of the main lobe as expected.

As is observed in Fig. \ref{pic_Pd_re}, in the case with $d = 0.5\lambda$,
the power of the arrived signal $P_{\mathrm{d}}$ decreases monotonically with $r_e$.
The concentration of the main lobe  along the  radial direction near the focused point is not be observed in this case,
since a small value of $d$ does not satisfy the constraint \eqref{SA_section3_B_eq3_1}.
This phenomenon results in severe interference for users located between the transmit array and the target user.
On the other hand, when the antenna spacing grows, 
it is observed from the case with $d = 10\lambda$ that 
the power of the arrived signal is concentrated at the focused point along the radial direction.
This is because that the phase variation induced by $r_e$ becomes the primary factor influencing the signal power.

Fig. \ref{pic_rlength_d} illustrates the variation of the main lobe length given in \eqref{SA_C2_eq2} as a function of the antenna spacing. Here, the sign of \( r_{\mathrm{length}} \) indicates whether the system exhibits the distance-focusing property: a positive value corresponds to the presence of the property, while a negative value indicates its absence. The vertical lines in the figure denote the minimum antenna spacing required to achieve distance focusing, which is observed to decrease as the number of antennas increases.

As discussed in the previous section, the strength of the distance-focusing capability is inversely proportional to the length of the main lobe. It can be observed that near the minimum antenna spacing, the main lobe length decreases rapidly with increasing \( d \), implying a sharp improvement in the distance-focusing capability. In addition, increasing the number of antennas also leads to a reduction in the main lobe length. However, comparable performance can be achieved more easily by increasing the antenna spacing \( d \). For instance, to attain a lobe length of 50 m, the product \( M \times N \) can be reduced from 2025 to 1225—a reduction of approximately 40\%—with only a \( 1.3\lambda \) increase in \( d \).

\begin{figure}[t]
	\centering
	\includegraphics[scale=0.50]{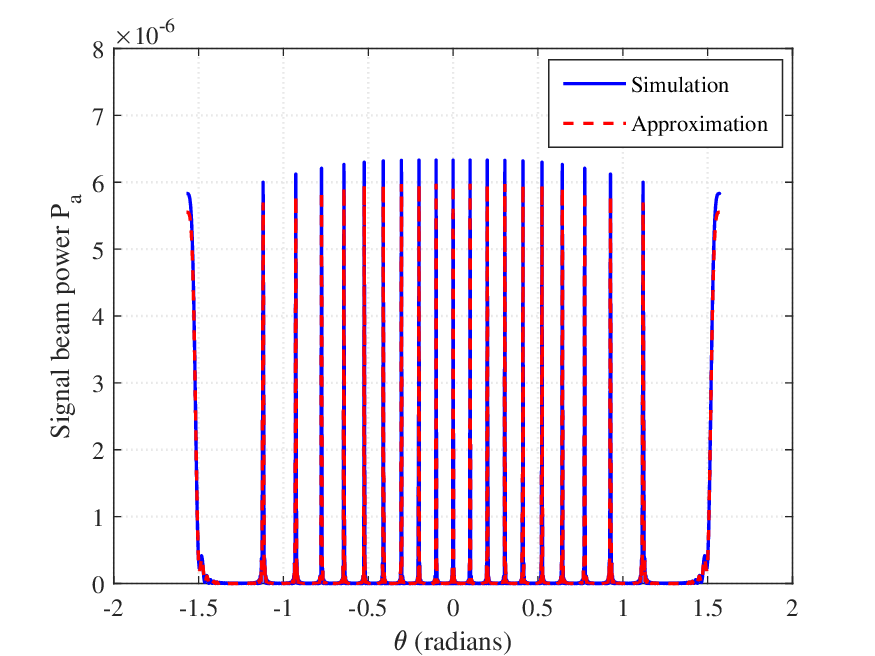}\\  
	\caption{Grating lobes in far-field ($r_0 = 100$ m).}\label{pic_lobe_far} 
\end{figure}

\begin{figure}[t]
	\centering
	\includegraphics[scale=0.50]{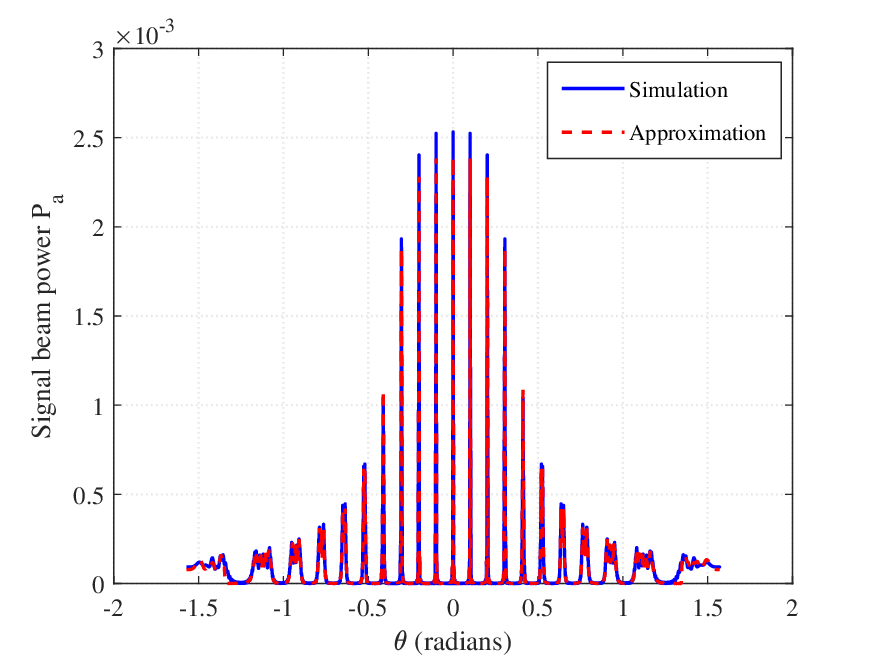}\\  
	\caption{Grating lobes in near-field ($r_0 = 5$ m).}\label{pic_lobe_near} 
\end{figure}

Fig. \ref{pic_lobe_far} illustrates the power distribution of grating lobes in the $\mathrm{XoZ}$ plane under far-field conditions at $r_0 = 100$ m. 
The antenna spacing is set as $d = 10 \lambda$.
The blue curve, labeled “Simulation,” is derived from \eqref{Pa_theta_sim}, while the red curve, labeled “Approximation,” is based on \eqref{Pa_theta_approx}. It can be observed that the two curves match closely across the entire range of elevation angles $\theta$, validating the accuracy of the closed-form approximation for grating lobe power given in \eqref{Pa_theta_approx}. Additionally, the results indicate that in the far-field case, the grating lobes exhibit power levels comparable to that of the main lobe, as expected.
This demonstrates that sparse UPAs lack grating lobe suppression capability in the far field.

Fig. \ref{pic_lobe_near} depicts the power distribution of grating lobes in the $\mathrm{XoZ}$ plane under near-field conditions at $r_0 = 5$ m.
A good agreement is observed between the simulation and the analytical approximation, which confirms the accuracy of the closed-form expression derived in \eqref{Pa_theta_approx}. 
Moreover, a clear suppression of the grating lobes is evident. 
In particular, the degree of suppression increases with the angular separation from the main lobe. 
This behavior can be attributed to the reduction in $\zeta_k$ defined in \eqref{zeta_definition}, which leads to a corresponding decrease in the ratio $\eta_k$.
These results demonstrate that grating lobes produced by sparse UPAs can be naturally suppressed in the near-field regime.

\begin{figure}[t]
	\centering
	\includegraphics[scale=0.50]{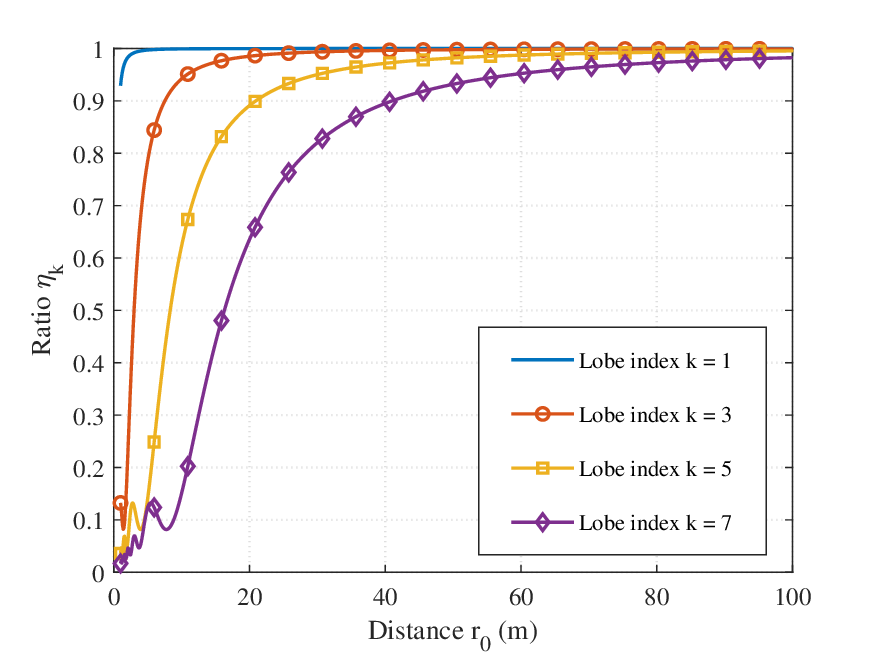}\\  
	\caption{Suppressing ratio $\eta_k$ versus $r_0$.}\label{pic_sup_ratio} 
\end{figure}

\begin{figure}[t]
	\centering
	\includegraphics[scale=0.50]{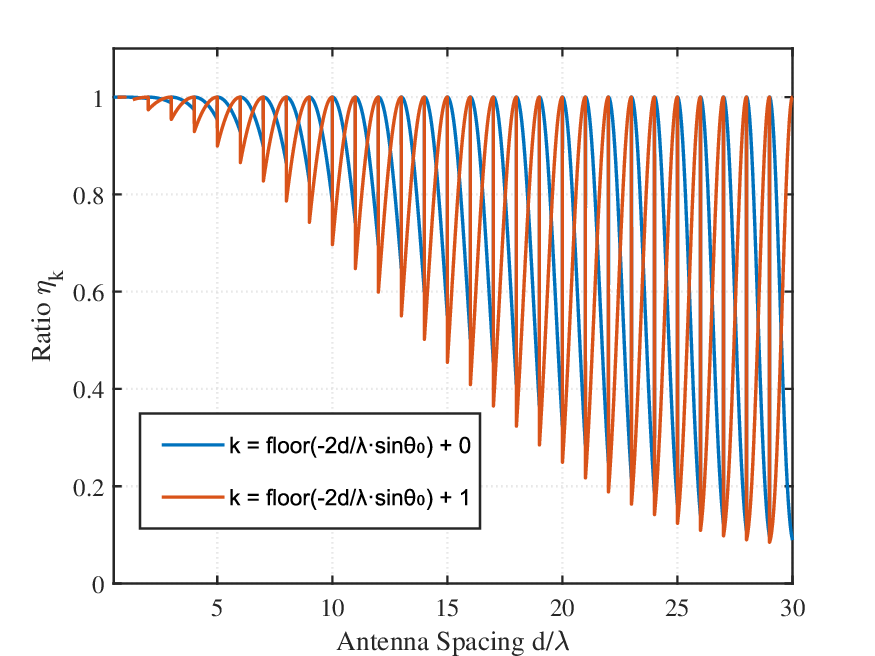}\\  
	\caption{Suppressing ratio $\eta_k$ of the strongest grating lobes versus $d$.}\label{pic_sup_ratio_d_k_highest} 
\end{figure}

\begin{figure}[t]
	\centering
	\includegraphics[scale=0.50]{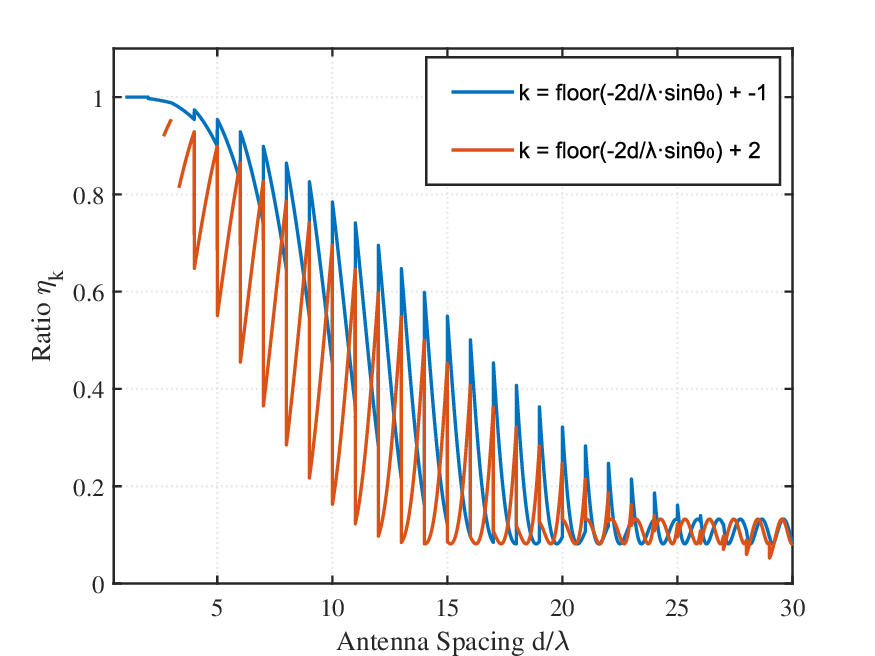}\\  
	\caption{Suppressing ratio $\eta_k$ of the sub-strongest grating lobes versus $d$.}\label{pic_sup_ratio_d_k_subhighest} 
\end{figure}

Fig. \ref{pic_sup_ratio} shows the variation of the suppressing ratio $\eta_k$ with respect to the focal distance $r_0$ for grating lobes with indices $\{1, 3, 5, 7\}$. 
It can be observed that as $r_0$ increases, the power levels of all grating lobes approach that of the main lobe. 
In contrast, when $r_0$ decreases, the suppressing ratios for grating lobes $\{3, 5, 7\}$ drop rapidly, indicating significant suppression in the near field. 
These findings are consistent with the results shown in Fig.~\ref{pic_lobe_far} and Fig.~\ref{pic_lobe_near}.

Fig. \ref{pic_sup_ratio_d_k_highest} examines the relationship between the antenna spacing \(d\) and the suppression ratio \(\eta_k\) of the most powerful grating lobes, whose indices are given by \eqref{index_k}. The elevation angle of the focused position is set to \(\theta_0 = -30^\circ\). It can be observed that as \(d\) increases, the power of these strongest grating lobes frequently approaches that of the main lobe, indicating that they cannot be effectively suppressed by increasing \(d\). In contrast, Fig. \ref{pic_sup_ratio_d_k_subhighest} shows the suppression ratio \(\eta_k\) for the sub-strongest grating lobes under the same variation of \(d\). Here, the results demonstrate that these lobes are effectively suppressed as \(d\) increases. These findings suggest that while increasing the antenna spacing can suppress most grating lobes, the particular ones identified in \eqref{index_k} remain largely unaffected. The simulation results are in strong agreement with the analysis provided in Corollary~\ref{Corollary_grating_lobe_index}.

\begin{figure}[t]
	\centering
	\includegraphics[scale=0.50]{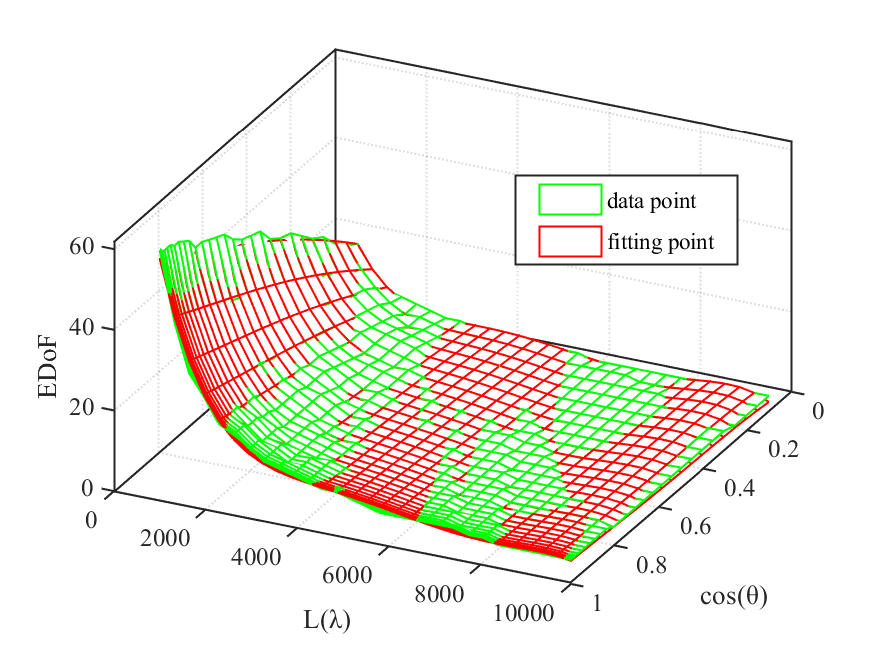}\\  
	\caption{Result of EDoF function fitting.}\label{SA_p11} 
\end{figure}

Fig.~\ref{SA_p11} compares the real data point for EDoF obtained by \eqref{SA_section4_B_eq1}  and the fitting point for EDoF obtained by \eqref{SA_section5_eq1}.
It is observed that the EDoF obtained by the closed-form expression $f_{\mathrm{EDoF}}\left( \theta ,r \right)$ well matches the real data point.
The normalized mean square error (NMSE) is 0.0068,
indicating a high accuracy of $f_{\mathrm{EDoF}}\left( \theta ,r \right)$.

\begin{figure}[t]
	\centering
	\includegraphics[scale=0.50]{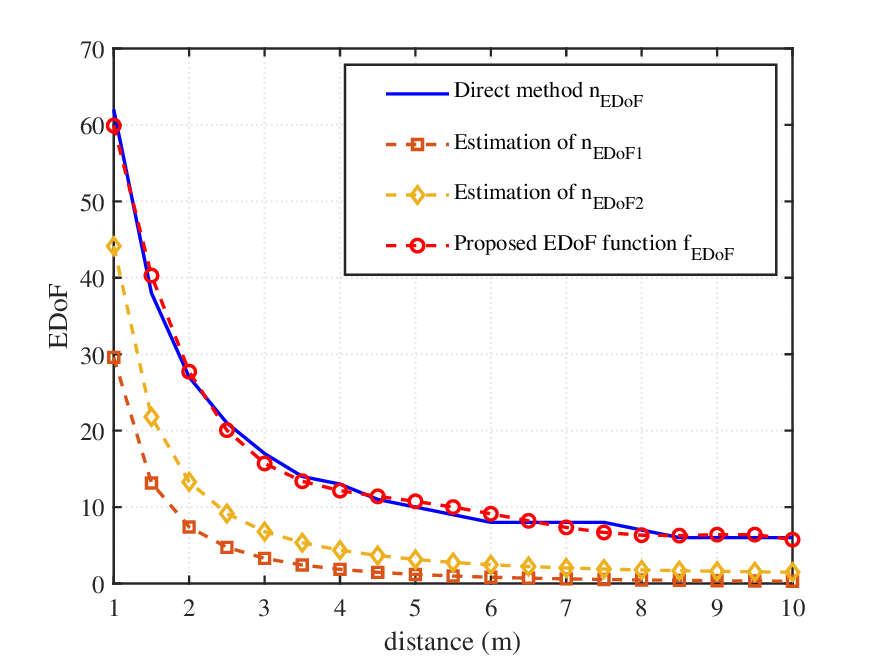}\\  
	\caption{Different EDoF estimation methods versus $r$.}\label{SA_p12} 
\end{figure}

\begin{figure}[t]
	\centering
	\includegraphics[scale=0.50]{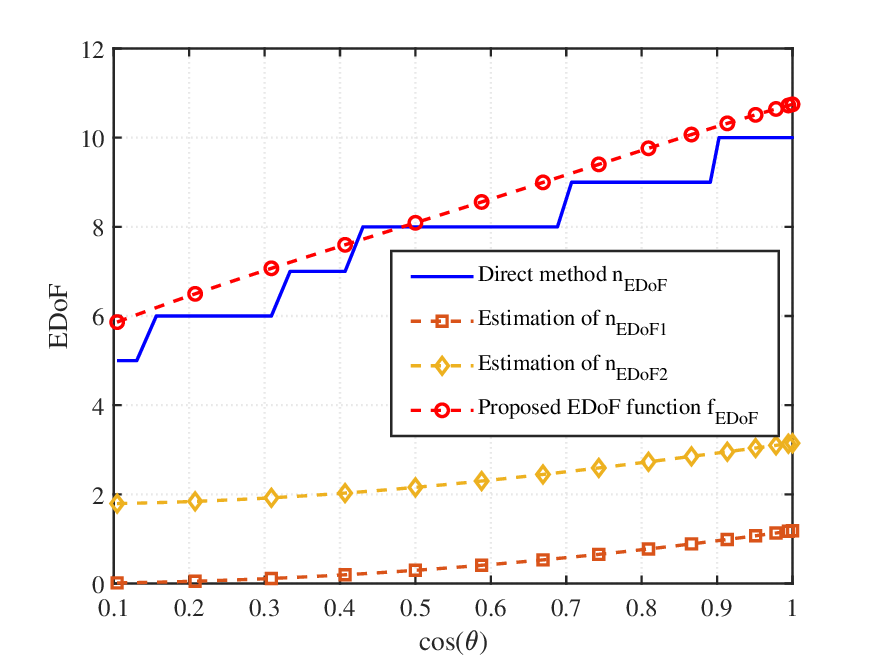}\\  
	\caption{Different EDoF estimation methods versus $\cos(\theta)$.}\label{SA_p13} 
\end{figure}

Fig.~\ref{SA_p12} and Fig.~\ref{SA_p13} compare $f_{\mathrm{EDoF}}\left( \theta ,r \right)$ with existing EDoF estimation methods as distance $r$ and angle $\theta$ vary, respectively.
The solid line of $n_{\mathrm{EDoF}}$ is obtained by \eqref{SA_section4_B_eq1},
which represents the real EDoF value.
The dashed line of $n_{\mathrm{EDoF1}}$ is obtained by \eqref{SA_section4_B_eq2},
the dashed line of $n_{\mathrm{EDoF2}}$ is based on \eqref{SA_section4_B_eq2} and 
the dashed line of $f_{\mathrm{EDoF}}$ represents our proposed method $f_{\mathrm{EDoF}}\left( \theta ,r \right)$.
Fig.~\ref{SA_p12} assumes $\cos(\theta) = 1$, 
while Fig.~\ref{SA_p13} is under the assumption of $r = 5$ m.
As is observed, the method in \eqref{SA_section4_B_eq2} has very limited accuracy, while 
the method in \eqref{SA_section4_B_eq2} provides a relatively better accuracy.
On the other hand, our proposed method provides a highly accurate estimation, significantly outperforming existing estimation methods.

\section{Conclusion}\label{SA_Section6}

We investigated near-field XL-MIMO systems with sparse UPAs.
Based on the Green's function-based channel model, 
we derived closed-form expressions for the signal beam power under the focused point $\mathbf{r}_0$,
and investigated two beneficial properties: the distance-focusing property and the grating lobe behavior in Section \ref{Sec_focusing} and Section \ref{Sec_suppression}, respectively.
Both of the properties can be enhanced by decreasing the focal distance $r_0$ or increasing the antenna spacing $d$. 
Furthermore, we introduced the constraint \eqref{SA_section4_C_eq1} on system parameters.
Based on that, we proposed Algorithm~\ref{SA_Algo2} to obtain an EDoF fitting function, which can estimate the increased EDoF of single-user XL-MIMO systems employing sparse UPAs with high accuracy and low computational complexity.
Finally, the numerical results verified the correctness of the main results of this paper.

\begin{figure*}[t]
	\setcounter{equation}{51}    
\begin{align}\label{app_Pd_expansion}
	& P_{\mathrm{d}}=\frac{P}{M^2N^2\left( 4\pi \left( r_0+r_e \right) \right) ^2}\left| \sum_{m=-\frac{M-1}{2}}^{\frac{M-1}{2}}{\sum_{n=-\frac{N-1}{2}}^{\frac{N-1}{2}}{\left( e^{j\frac{2\pi}{\lambda}r_{m,n}\left( r_0,\theta _0,\phi _0 \right)}e^{-j\frac{2\pi}{\lambda}r_{m,n}\left( r,\theta _0,\phi _0 \right)} \right)}} \right|^2
	\nonumber \\
    & =\frac{P}{M^2N^2\left( 4\pi \left( r_0+r_e \right) \right) ^2}\left| \sum_{m=-\frac{M-1}{2}}^{\frac{M-1}{2}}{e^{j\frac{\pi}{\lambda}\frac{r_e}{r_0\left( r_0+r_e \right)}\left( \cos ^2\theta _0+\sin ^2\theta _0\sin ^2\phi _0 \right) d^{2}m^2}} \right|^2\left| \sum_{n=-\frac{N-1}{2}}^{\frac{N-1}{2}}{e^{j\frac{\pi}{\lambda}\frac{r_e}{r_0\left( r_0+r_e \right)}\left( \cos ^2\theta _0+\sin ^2\theta _0\cos ^2\phi _0 \right) d^{2}n^2}} \right|^2
\end{align}
	\hrulefill
	\setcounter{equation}{55}    
	\begin{align}\label{app_varepsilon_x_approx}
		& \varepsilon _x=\int_{-\frac{M-1}{2}}^{\frac{M-1}{2}}{\cos \left( \frac{\pi}{\lambda}\frac{r_ed^2x^2}{r_0\left( r_0+r_e \right)}\left( \cos ^2\theta _0+\sin ^2\theta _0\sin ^2\phi _0 \right) \right) dx} + j\int_{-\frac{M-1}{2}}^{\frac{M-1}{2}}{\sin \left( \frac{\pi}{\lambda}\frac{r_ed^2x^2}{r_0\left( r_0+r_e \right)}\left( \cos ^2\theta _0+\sin ^2\theta _0\sin ^2\phi _0 \right) \right) dx}   \nonumber \\
		& \xlongequal{t=\tau _x\mu x, \mu =d\sqrt{\frac{2}{\lambda}\frac{r_e}{r_0\left( r_0+r_e \right)}}, \frac{r_e}{r_0+r_e}>0}\frac{1}{\tau _x\mu}\int_{-\frac{M-1}{2}\tau _x\mu}^{\frac{M-1}{2}\tau _x\mu}{\cos \left( \frac{\pi}{2}t^2 \right) dt} + j\frac{1}{\tau _x\mu}\int_{-\frac{M-1}{2}\tau _x\mu}^{\frac{M-1}{2}\tau _x\mu}{\sin \left( \frac{\pi}{2}t^2 \right) dt}  \nonumber \\
	    & =\frac{1}{\tau _x\mu}\left[ C\left( b_M \right) -C\left( -b_M \right) +j\left( S\left( b_M \right) -S\left( -b_M \right) \right) \right] =\frac{2}{\tau _x\mu}\left[ C\left( b_M \right) +jS\left( b_M \right) \right] =\left( M-1 \right) \frac{1}{b_M}\left[ C\left( b_M \right) +jS\left( b_M \right) \right] 
	\end{align}
	\hrulefill

	\setcounter{equation}{50}
\end{figure*}

\begin{appendices}

\section{Proof for Theorem \ref{Theorem_distance}}\label{app_distance}

When the XL-MIMO array focuses on $\mathbf{r}_0$, 
the power $P_{\mathrm{d}}$ of the signal $f_{\mathrm{d}}$ arrived at $\mathbf{r}_{\mathrm{d}}$ can be obtained from \eqref{fd_expression} as \vspace{-0.1cm}
\begin{equation}\label{app_Pd_sim}\vspace{-0.1cm} 
	P_{\mathrm{d}}=\frac{P}{MN}\left| \sum_{m=-\frac{M-1}{2}}^{\frac{M-1}{2}}{\sum_{n=-\frac{N-1}{2}}^{\frac{N-1}{2}}{\left( w_{m,n}^{0} \right) ^{\ast}h_{m,n}^{\mathrm{d}}}} \right|^2 .
\end{equation}
Substituting \eqref{diatance_rmn_approx}, \eqref{w0_expression} and \eqref{hd_expression} into \eqref{app_Pd_sim},
we can calculate $P_{\mathrm{d}}$ as \eqref{app_Pd_expansion} at the top of the next page. \addtocounter{equation}{1}

To further calculate $P_{\mathrm{d}}$, we define
\begin{equation}
	\varepsilon _x= \sum_{m=-\frac{M-1}{2}}^{\frac{M-1}{2}}{e^{j\frac{\pi}{\lambda}\frac{r_e}{r_0\left( r_0+r_e \right)}\left( \cos ^2\theta _0+\sin ^2\theta _0\sin ^2\phi _0 \right) d^{2}m^2}} ,
\end{equation}
\begin{equation}
	\varepsilon _y= \sum_{n=-\frac{N-1}{2}}^{\frac{N-1}{2}}{e^{j\frac{\pi}{\lambda}\frac{r_e}{r_0\left( r_0+r_e \right)}\left( \cos ^2\theta _0+\sin ^2\theta _0\cos ^2\phi _0 \right) d^{2}n^2}} .
\end{equation}
Then, we can approximate $\varepsilon_x $ and $\varepsilon_y $ using Euler-Maclaurin formula, 
which is a method for approximating summation terms by transforming the summation into an integral, thereby simplifying the calculation \cite{1999Advanced}.

For $\varepsilon_x $, we have \vspace{-0.1cm}
\begin{equation}\label{SA_App_B_eq9}\vspace{-0.1cm} 
	\varepsilon _x\approx \int_{-\frac{M-1}{2}}^{\frac{M-1}{2}}{e^{j\frac{\pi}{\lambda}\frac{r_e}{r_0\left( r_0+r_e \right)}\left( \cos ^2\theta _0+\sin ^2\theta _0\sin ^2\phi _0 \right) d^{2}x^2}dx},
\end{equation}
which can be further expressed as \eqref{app_varepsilon_x_approx} at the top of the next page, 
where functions $C(\cdot )$ and $S(\cdot )$ are Fresnel integrals, respectively given by \eqref{Theorem1_CandS}, 
Parameters $b_M$, $\tau_x$ and $\mu$ are given by \eqref{Theorem1_parameters}.
\addtocounter{equation}{1}

It is noted that Eq. \eqref{app_varepsilon_x_approx} analyzes the case of $\frac{r_e}{r_0+r_e}>0$. For the case of $\frac{r_e}{r_0+r_e}<0$, though $\varepsilon _x$ becomes conjugate, its norm version in \eqref{app_Pd_expansion} remains the same. 
Additionally, for $\frac{r_e}{r_0+r_e}=0$, we have $\varepsilon _x=M-1$, since 
$
C\left( x \right) =x-\frac{\pi ^2x^5}{40}+\mathcal{O} \left( x^9 \right)
$
and
$
S\left( x \right) =\frac{\pi x^3}{6}-\frac{\pi ^3x^7}{336}+\mathcal{O} \left( x^{11} \right) 
$ as $x\rightarrow 0$.

Similarly, the same method can be used to derive the approximation of $\varepsilon_x $, 
which is expressed as
\begin{equation}\label{app_varepsilon_y_approx}
	\varepsilon _y=\left( N-1 \right) \frac{1}{b_N}\left[ C\left( b_N \right) +jS\left( b_N \right) \right] ,
\end{equation}
where parameters $b_N$, $\tau_y$ and $\mu$ are given by \eqref{Theorem1_parameters}.

Substituting \eqref{app_varepsilon_x_approx} and \eqref{app_varepsilon_y_approx} into \eqref{app_Pd_expansion},
we arrive at \eqref{Theorem_distance_Pd_approx},
and thus complete the proof for Theorem \ref{Theorem_distance}.

\begin{figure*}[t]
	\setcounter{equation}{62}    
	\begin{align}\label{app_Pa_expansion}
		& P_{\mathrm{a}}=\frac{P}{M^2N^2\left( 4\pi r_0 \right) ^2}\left| \sum_{m=-\frac{M-1}{2}}^{\frac{M-1}{2}}{\sum_{n=-\frac{N-1}{2}}^{\frac{N-1}{2}}{\left( e^{j\frac{2\pi}{\lambda}r_{m,n}\left( r_0,\theta _0,\phi _0 \right)}e^{-j\frac{2\pi}{\lambda}r_{m,n}\left( r_0,\theta ,\phi \right)} \right)}} \right|^2
		\nonumber \\
		& =\frac{P}{M^2N^2\left( 4\pi r_0 \right) ^2}\left| \sum_{m=-\frac{M-1}{2}}^{\frac{M-1}{2}}{e^{j2\pi \left( m\frac{d}{\lambda}\left( \sin \theta _0\cos \phi _0-\sin \theta \cos \phi \right) +m^2\frac{d^2}{2r_0\lambda}\left( \cos ^2\theta -\cos ^2\theta _0+\sin ^2\theta \sin ^2\phi -\sin ^2\theta _0\sin ^2\phi _0 \right) \right)}} \right|^2	
		\nonumber \\
		& \times \left| \sum_{n=-\frac{N-1}{2}}^{\frac{N-1}{2}}{e^{j2\pi \left( n\frac{d}{\lambda}\left( \sin \theta _0\sin \phi _0-\sin \theta \sin \phi \right) +n^2\frac{d^2}{2r_0\lambda}\left( \cos ^2\theta -\cos ^2\theta _0+\sin ^2\theta \cos ^2\phi -\sin ^2\theta _0\cos ^2\phi _0 \right) \right)}} \right|^2
		\nonumber \\
		& \overset{\left( a \right)}{=}\frac{P}{M^2N^2\left( 4\pi r_0 \right) ^2}\left| \sum_{m=-\frac{M-1}{2}}^{\frac{M-1}{2}}{e^{j2\pi \left( b_xm+a_xm^2 \right)}} \right|^2\left| \sum_{n=-\frac{N-1}{2}}^{\frac{N-1}{2}}{e^{j2\pi \left( b_yn+a_yn^2 \right)}} \right|^2
		\nonumber \\
		& =\frac{P}{M^2N^2\left( 4\pi r_0 \right) ^2}\left| \sum_{m=-\frac{M-1}{2}}^{\frac{M-1}{2}}{e^{-j\frac{b_{x}^{2}\pi}{2a_x}}e^{j2\pi a_x\left( m+\frac{b_x}{2a_x} \right) ^2}} \right|^2\left| \sum_{n=-\frac{N-1}{2}}^{\frac{N-1}{2}}{e^{-j\frac{b_{y}^{2}\pi}{2a_y}}e^{j2\pi a_y\left( m+\frac{b_y}{2a_y} \right) ^2}} \right|^2
		\nonumber \\
		& \overset{\left( b \right)}{=}\frac{P}{M^2N^2\left( 4\pi r_0 \right) ^2}\left| \sum_{m=-\frac{M-1}{2}}^{\frac{M-1}{2}}{e^{j2\pi a_x\left( m+\frac{b_x}{2a_x} \right) ^2}} \right|^2\left| \sum_{n=-\frac{N-1}{2}}^{\frac{N-1}{2}}{e^{j2\pi a_y\left( m+\frac{b_y}{2a_y} \right) ^2}} \right|^2\overset{\left( c \right)}{=}\frac{P}{M^2N^2\left( 4\pi r_0 \right) ^2}\left| \epsilon _x\epsilon _y \right|^2
	\end{align}
	\hrulefill

	\setcounter{equation}{57}
\end{figure*}

\section{Proof for Theorem \ref{SA_C1}}\label{app_rho_d}

According to the definition of $\mu$ in \eqref{Theorem1_parameters}, 
it is readily observed that $\mu$ has the range of $\left[ 0,+\infty \right) $.
Then, by differentiating $\rho _{\mathrm{d}}(\mu)$, we investigate the monotonicity of $\rho _{\mathrm{d}}$.

The derivative of $\rho _{\mathrm{d}}(\mu)$ is calculated as
\begin{equation*}
	\left( \rho _{\mathrm{d}} \right) _{\mu}^{\prime}=\frac{\left( M-1 \right) \tau _x\left( C^2(b_N)+S^2(b_N) \right)}{b_{N}^{2}}g\left( b_M \right)
\end{equation*}
\begin{equation}\label{app_derivative_pho_d}
	 +\frac{\left( N-1 \right) \tau _y\left( C^2(b_M)+S^2(b_M) \right)}{b_{M}^{2}}g\left( b_N \right)   ,
\end{equation}
where function $g(x), x>0$ is defined as
\begin{equation*}
	g\left( x \right) = -\frac{1}{x^3}[C^2(x)+S^2(x)]
\end{equation*}
\begin{equation}\label{app_gx}
	+ \frac{1}{x^2}\left[ C(x)\cos \left( \frac{\pi x^2}{2} \right) +S(x)\sin \left( \frac{\pi x^2}{2} \right) \right] .
\end{equation}

It is noted from \eqref{app_gx} that the sign of $g(x)$ depends on the following two terms 
\begin{equation}\label{SA_App_C_eq8}
	\hspace{-2.6cm}
	\mathrm{term}1=-\frac{1}{x^3}\left( C^2\left( x \right) +S^2\left( x \right) \right)  ,
\end{equation}
\begin{equation}\label{SA_App_C_eq9}
	\mathrm{term}2=\frac{1}{x^2}\left[ C(x)\cos \left( \frac{\pi x^2}{2} \right) +S(x)\sin \left( \frac{\pi x^2}{2} \right) \right]  .
\end{equation}
When $x$ is close to 0, $\mathrm{term}1$ dominates the sign of $g(x)$, since the factor \( -\frac{1}{x^3} \) causes \( g(x) \) to become a negative value with large magnitude.
When $b$ becomes large, the absolute value of $\mathrm{term}1$ drops off, and thus $\mathrm{term}2$ dominates the sign of $g(x)$, 
which is a function of $x$ with finite oscillations, alternating between positive and negative values.

Furthermore, since $b_M$ and $b_N$ are proportional to $\mu$, the behavior of $g(x)$ implies that for small $\mu$, the derivative $\left( \rho_{\mathrm{d}} \right)_{\mu}^{\prime}$ in \eqref{app_derivative_pho_d} is negative.  
As $\mu$ becomes large, $\left( \rho_{\mathrm{d}} \right)_{\mu}^{\prime}$ oscillates with bounded amplitude between positive and negative values.

These properties imply that over the domain $\mu >0$,
$\rho _{\mathrm{d}}(\mu)$ initially decreases as $\mu$ increases, and subsequently fluctuates within a limited range.
This completes the proof of Theorem~\ref{SA_C1}.

\begin{figure*}[t]
	\setcounter{equation}{67}
	\begin{align}\label{Pa_theta_angular_location}
		&  P_{\mathrm{a},\theta}\approx P_{\mathrm{a},\theta}^{\prime}=\frac{P}{M^2\left( 4\pi r_0 \right) ^2}\left| \sum_{m=-\frac{M-1}{2}}^{\frac{M-1}{2}}{e^{mj2\pi \frac{d}{\lambda}\left( \sin \theta -\sin \theta _0 \right)}} \right|^2 
		\nonumber \\
		&  =\frac{P}{M^2\left( 4\pi r_0 \right) ^2}\left| \frac{e^{-\frac{M-1}{2}j2\pi \frac{d}{\lambda}\left( \sin \theta -\sin \theta _0 \right)}\left( 1-e^{Mj2\pi \frac{d}{\lambda}\left( \sin \theta -\sin \theta _0 \right)} \right)}{1-e^{j2\pi \frac{d}{\lambda}\left( \sin \theta -\sin \theta _0 \right)}} \right|^2
		\nonumber \\
		&  =\frac{P}{M^2\left( 4\pi r_0 \right) ^2}\left| \frac{e^{-\left( M-1 \right) j\pi \frac{d}{\lambda}\left( \sin \theta -\sin \theta _0 \right)}e^{Mj\pi \frac{d}{\lambda}\left( \sin \theta -\sin \theta _0 \right)}\left( e^{-Mj\pi \frac{d}{\lambda}\left( \sin \theta -\sin \theta _0 \right)}-e^{Mj\pi \frac{d}{\lambda}\left( \sin \theta -\sin \theta _0 \right)} \right)}{e^{j\pi \frac{d}{\lambda}\left( \sin \theta -\sin \theta _0 \right)}\left( e^{-j\pi \frac{d}{\lambda}\left( \sin \theta -\sin \theta _0 \right)}-e^{j\pi \frac{d}{\lambda}\left( \sin \theta -\sin \theta _0 \right)} \right)} \right|^2
		\nonumber \\
		&  =\frac{P}{M^2\left( 4\pi r_0 \right) ^2}\left| \frac{\sin \left( M\pi \frac{d}{\lambda}\left( \sin \theta -\sin \theta _0 \right) \right)}{\sin \left( \pi \frac{d}{\lambda}\left( \sin \theta -\sin \theta _0 \right) \right)} \right|^2=\frac{P}{M^2\left( 4\pi r_0 \right) ^2}\left| \frac{\sin \left( M\nu \right)}{\sin \left( \nu \right)} \right|^2 
	\end{align}
	\hrulefill
	
		\setcounter{equation}{71}
	\begin{equation}\label{Pa_theta_distance_kappa}
		\kappa =\frac{\partial}{\partial m}\left( 2\pi \left( m\frac{d}{\lambda}\left( \sin \theta -\sin \theta _0 \right) +m^2\frac{d^2}{2r_0\lambda}\left( \cos ^2\theta _0-\cos ^2\theta \right) \right) \right) =\frac{d}{\lambda}\left( \sin \theta -\sin \theta _0 \right) +m\frac{d^2}{r_0\lambda}\left( \cos ^2\theta _0-\cos ^2\theta \right) 
	\end{equation}
	\hrulefill
	\begin{align}\label{Pa_theta_sim_k}
		& P_{\mathrm{a},\theta}=\frac{P}{M^2\left( 4\pi r_0 \right) ^2}\left| \sum_{m=-\frac{M-1}{2}}^{\frac{M-1}{2}}{e^{j2\pi \left( m\frac{d}{\lambda}\left( \sin \theta -\sin \theta _{k}^{\prime}+\frac{k\lambda}{d} \right) +m^2\frac{d^2}{2r_0\lambda}\left( \cos ^2\theta _0-\cos ^2\theta \right) \right)}} \right|^2  \nonumber \\
		& \overset{\left( a \right)}{=}\frac{P}{M^2\left( 4\pi r_0 \right) ^2}\left| \sum_{m=-\frac{M-1}{2}}^{\frac{M-1}{2}}{e^{j2\pi \left( m\frac{d}{\lambda}\left( \sin \theta -\sin \theta _{k}^{\prime} \right) +m^2\frac{d^2}{2r_0\lambda}\left( \cos ^2\theta _0-\cos ^2\theta \right) \right)}} \right|^2
	\end{align}
	\hrulefill
	\setcounter{equation}{61}
\end{figure*}

\section{Proof for Theorem \ref{Theorem_angle}}\label{app_angle}

When the XL-MIMO array focuses on $\mathbf{r}_0$, 
the power $P_{\mathrm{a}}$ of the signal $f_{\mathrm{a}}$ arrived at $\mathbf{r}_{\mathrm{a}}$ can be obtained from \eqref{fa_expression} as \vspace{-0.1cm}
\begin{equation}\label{app_Pa_sim}\vspace{-0.1cm} 
	P_{\mathrm{a}}=\frac{P}{MN}\left| \sum_{m=-\frac{M-1}{2}}^{\frac{M-1}{2}}{\sum_{n=-\frac{N-1}{2}}^{\frac{N-1}{2}}{\left( w_{m,n}^{0} \right) ^{\ast}h_{m,n}^{\mathrm{a}}}} \right|^2 .
\end{equation}
Substituting \eqref{diatance_rmn_approx}, \eqref{w0_expression} and \eqref{ha_expression} into \eqref{app_Pa_sim},
we can calculate $P_{\mathrm{a}}$ as \eqref{app_Pa_expansion} at the top of the previous page. \addtocounter{equation}{1}
The step $(a)$ is based on \eqref{Theorem_angle_abx} and \eqref{Theorem_angle_aby}.
The step $(b)$ is due to the fact that $\left| e^{-j\frac{b^2\pi}{2a}} \right|=1$.
The step $(c)$ is obtained by defining
\begin{equation}
	\epsilon _x=\sum_{m=-\frac{M-1}{2}}^{\frac{M-1}{2}}{e^{j2\pi a_x\left( m+\frac{b_x}{2a_x} \right) ^2}},
\end{equation}
\begin{equation}
	\epsilon _y=\sum_{n=-\frac{N-1}{2}}^{\frac{N-1}{2}}{e^{j2\pi a_y\left( m+\frac{b_y}{2a_y} \right) ^2}}.
\end{equation}

Utilizing the similar approximation method for $\varepsilon _x$ in \eqref{app_varepsilon_x_approx}, 
the approximations of $\epsilon _x$ and $\epsilon _y$ can be derived respectively as
\begin{equation}\label{app_epsilon_x_approx}
	\epsilon _x\approx \frac{1}{\sqrt{4\left| a_x \right|}}\left( \left( C\left( u_{1,x} \right) \! + \! C\left( u_{2,x} \right) \right)  \! + \! j \! \left( S\left( u_{1,x} \right) \!  + \! S\left( u_{2,x} \right) \right) \right) ,
\end{equation}
\begin{equation}\label{app_epsilon_y_approx}
	\epsilon _y\approx \frac{1}{\sqrt{4\left| a_y \right|}}\left( \left( C\left( u_{1,y} \right)  \! + \! C\left( u_{2,y} \right) \right)  \! + \! j \! \left( S\left( u_{1,y} \right)  \! + \! S\left( u_{2,y} \right) \right) \right) ,
\end{equation}
where $u_{1,x}$, $u_{2,x}$, $u_{1,y}$ and $u_{2,y}$ are given by \eqref{Theorem_angle_ux} and \eqref{Theorem_angle_uy}.

Substituting \eqref{app_epsilon_x_approx} and \eqref{app_epsilon_y_approx} into \eqref{app_Pa_expansion},
we arrive at \eqref{Theorem_angle_Pa_approx},
and thus complete the proof for Theorem \ref{Theorem_angle}.

\section{Proof for Corollary \ref{Corollary_grating_lobe_location}}\label{app_angle_location}

It is well-established that both the main lobe and grating lobes exhibit substantial power. To examine the angular distribution of high-power regions, we simplify expression \eqref{Pa_theta_sim} into the form given in \eqref{Pa_theta_angular_location} at the top of this page, by neglecting a second-order phase term that is significantly smaller than the first-order term.  \addtocounter{equation}{1}
Here, $\nu$ is defined as
\begin{equation}
	\nu =\pi \frac{d}{\lambda}\left( \sin \theta -\sin \theta _0 \right).
\end{equation}
This approximation preserves the angular locations of the lobes, as the neglected term has negligible influence on the position of the power maxima.

Observing that \( P_{\mathrm{a},\theta}^{\prime}(\nu) \) in \eqref{Pa_theta_angular_location} exhibits a period of \(\pi\), the angular relationship between the main lobe at \(\theta = \theta_0\) and the grating lobes at \(\theta = \theta_{k}^{\prime}\) can be concisely expressed as
\begin{align}
	&
	\nu \left( \theta =\theta _0 \right) =\nu \prime\left( \theta =\theta _{k}^{\prime} \right) +k\pi 
	\nonumber \\
	&  \Rightarrow \pi \frac{d}{\lambda}\left( \sin \theta _0-\sin \theta _0 \right) =\pi \frac{d}{\lambda}\left( \sin \theta _{k}^{\prime}-\sin \theta _0 \right) +k\pi 
	\nonumber \\
	&  \Rightarrow \sin \theta _{k}^{\prime}=\sin \theta _0+\frac{k\lambda}{d}
	\nonumber \\
	&  \Rightarrow \theta _{k}^{\prime}=\mathrm{arc}\sin \left( \sin \theta _0+\frac{k\lambda}{d} \right), \;\;\;  k=0, \pm 1,\pm 2,... .
\end{align}
Moreover, the range limitation of $\mathrm{arc}\sin (\cdot)$ further requires
\begin{equation}
	\sin \theta _0+\frac{k\lambda}{d}\in \left[ -1,1 \right],
\end{equation}
and thus the final range of integer $k$ is obtained as \eqref{Corollary_k_range}.
This completes the proof.

\section{Proof for Theorem \ref{Theorem_grating_lobe_intensity}}\label{app_angle_intensity}

To approximate the summation term in \eqref{Pa_theta_sim} as an integral using the Euler–Maclaurin formula, the phase of the exponential function must vary slowly between successive integers \(m\). This requires the derivative of the phase \(\kappa\) with respect to \(m\), given by \eqref{Pa_theta_distance_kappa} at the top of the previous page, \addtocounter{equation}{1} to be sufficiently small. The second term in \eqref{Pa_theta_distance_kappa} is small enough due to the \(1/r\) factor, whereas the first term may become excessively large when \(\theta\) deviates significantly from \(\theta_0\), particularly in the case of a sparse array.

To mitigate this issue, when $\theta$ is close to $\theta _{k}^{\prime}$, we apply the relation \eqref{Corollary_theta_k} to replace \(\theta_0\) in \eqref{Pa_theta_sim}, leading to \eqref{Pa_theta_sim_k} at the top of the previous page. \addtocounter{equation}{1} Step \((a)\) follows from the fact that \(\left| e^{j2\pi mk} \right| = 1\). This substitution effectively reduces the derivative of the phase in the form given by \eqref{Pa_theta_sim_k}. Subsequently, a procedure analogous to that in \eqref{SA_App_B_eq9} can be applied, yielding the closed-form expression presented in \eqref{Pa_theta_approx}.


\end{appendices}

\bibliographystyle{IEEEtran}
\bibliography{IEEEabrv,Reference}

\end{document}